**Femicide Laws, Unilateral Divorce, and Abortion Decriminalization Fail to Stop Women from Being Killed in Mexico**

**Roxana Gutiérrez-Romero**[♣]

July 2025

## Abstract

This paper evaluates the effectiveness of femicide laws in combating gender-based killings of women, a major cause of premature female mortality. Focusing on Mexico, a pioneer in adopting such legislation, the paper exploits variations in the enactment of femicide laws and prison sentences across states. Using the difference-in-differences estimator, the analysis reveals femicide laws have not impacted femicides, homicides, disappearances, or suicides of women. Results remain robust when considering differences in prison sentencing, states introducing unilateral divorce, equitable divorce asset compensation, or decriminalizing abortion. Findings also hold with synthetic matching, suggesting laws are insufficient to combat gender-based violence in contexts of impunity.

---

[♣] Queen Mary University of London, London r.gutierrez@qmul.ac.uk

I acknowledge funding from the Global Challenges Research Fund (GCRF) [RE-CL-2021-01] and the Impact Fellowship at the School of Business and Management (SBM) at Queen Mary University of London.

## 1. Introduction

Femicides—the hate killings of girls and women because of their gender—make up a small fraction of total homicides, but their rise is deeply alarming (Koppa and Messing 2021). For instance, in 2022, the world saw 243 women murdered daily, the highest annual figure in two decades (UNODC 2022). Nearly 40% of these victims were killed by intimate partners or ex-partners, making femicide one of the leading causes of premature female deaths (Corradi 2021; WHO 2021). While gender-based violence is a global issue, the surge in femicides is especially severe in developing regions, notably Latin America and the Caribbean (Small Arms Survey 2018). This region has experienced the fastest growth in such crimes, prompting 18 out of 33 countries to enact femicide laws aimed at confronting this public health crisis. These laws typically categorize femicides as hate crimes and impose stricter penalties than those for conventional homicides (García-Del Moral and Neumann 2019). Recently, some European nations, including Belgium, Croatia, Cyprus, and Malta have also adopted femicide laws, with many others considering similar measures. However, despite public debate, the effectiveness of these laws remains uncertain, as no comprehensive impact evaluation has been conducted to date.[1]

This paper is the first to evaluate the impact of femicide laws on lethal violence against women, considering outcomes such as femicides, homicides, suicides, and disappearances. I focus on Mexico, which was one of the first countries to recognize gender-based homicide as an aggravating factor and to subsequently enact femicide legislation nationwide. Between 2010 and 2019, all 32 states gradually adopted femicide laws, though with significant variation in prison sentencing severity. What is unique to Mexico, is that during this period, several states also introduced other legal reforms affecting family and women's rights, including unilateral divorce laws (26 states), equitable asset distribution upon divorce (23 states), and abortion decriminalization (two states). These policy changes provide important insights into whether other concurrent legal reforms may enhance or undermine the effectiveness of femicide laws.

For instance, theoretically, unilateral divorce laws may contribute to reducing homicides by enabling women to leave abusive relationships more easily, as seen in high-income countries (Stevenson and Wolfers 2006). However, in Mexico, most states that enacted unilateral divorce also introduced a provision requiring up to 50% of marital assets to be transferred to the spouse responsible for household and childcare duties when the couple had married under a separate property regime. While most marriages follow the community property regime, which already ensures equitable asset division, about 15% of marriages opt for the separate property regime. For this minority, the reform represents a significant departure from previous norms, where divorced individuals kept only the assets registered in their name. Intended to promote economic fairness, this provision may provoke a violent backlash from partners facing divorce who could perceive it as an unfair financial burden. Similarly, while the decriminalization of abortion expands women's reproductive freedom, it could trigger violence from partners who oppose or seek to control abortion decisions (Pallitto, Campbell, and O'campo 2005; Muratori 2021). When combined with femicide laws imposing harsher penalties, these concurrent legal changes may deter some forms of violence but could also incentivize perpetrators to adapt, such as concealing crimes—through disappearances or staged suicides—to evade punishment.

[1] Numerous studies have monitored the trend of femicides before and since the implementation of such femicide laws (Koppa and Messing 2021; Lombard 2018; García-Del Moral and Neumann 2019; Sarmiento et al. 2014). Yet, to the best of my knowledge, no rigorous impact evaluation of the effectiveness of the new legislation has been conducted.

To evaluate the impact of femicide laws along with associated policy changes I exploit the gradual adoption of this legislation across Mexican states using the staggered difference-in-differences estimator proposed by de Chaisemartin and D'Haultfoeuille (2022). I begin by estimating the impact of femicide laws—analyzing states that implemented exclusively such legislation compared to those that did not. Then, I take advantage of the great variability in the timing of enactment of other laws, to derive causal estimates of the impact of femicide laws when complemented by other legislative changes, such as unilateral divorce reforms, new equitable divorce asset compensation, or the decriminalization of abortion.

Another key component of the legislative changes is the difference in prison sentences for femicides and homicides among states and over time. The staggered difference-in-differences estimator used allows me to assess whether tougher sentencing for femicide had a differential impact compared to states using more lenient penalties. Additionally, I examine whether the differences between femicide-specific and general homicide-specific sentencing had any impact on lethal violence against women.

A major challenge in evaluating the impact of femicide laws is the lack of reliable official data. By law, femicides in Mexico are defined as homicides of women exhibiting one or more of the following seven characteristics: sexual abuse, antecedents of violence or threats, public disposal of the body, degrading injuries inflicted before or after death, abduction, or having a close relationship with the perpetrator. Despite having this legal definition, official femicide statistics were not compiled until 2015—five years after the first-ever femicide state laws were enacted, and after such laws had already been adopted by nearly all states. Even after data collection began, systematic misclassification has persisted. My estimates indicate that at least 60% of femicides are officially recorded as homicides in judicial investigations partly due to flawed investigative protocols and chronic underfunding (McGinnis, Rodríguez Ferreira, and Shirk 2022; Amnesty International 2021).[2]

To address the lack of reliable official femicide data, I use four indicators that provide a comprehensive view of lethal violence against women, including the period before femicide laws were enacted. The first three indicators are the monthly municipal-level counts of female homicides (regardless of the perpetrator's motive or intent, excluding accidental deaths), likely femicides, and suicides. All these statistics are obtained from official death certificates. These certificates, by law, are issued by forensic medical services prior to judicial investigations, largely focused on medical forensic evidence.[3] The death certificates do not categorize femicides. However, they classify deaths as accidental, homicidal, suicidal, or due to natural causes, including important details such as the cause of death, indications of sexual violence, and, in some instances, the identity of the suspected perpetrator as reported by the victim's next of kin. I use this information to identify likely femicides, even during periods when states had not yet adopted a legal definition for femicide. As a fourth indicator, I include the official monthly municipal counts of girls and women reported as disappeared, likely for being victims of a crime.

A potential concern in evaluating femicide laws is determining whether their staggered adoption across Mexican states was an exogenous policy shift or driven by pre-existing violence trends in certain states. The 1990s murders in Ciudad Juárez, on Mexico's northern border, undoubtedly brought global attention to the country's crisis of gender-based violence. However, the subsequent enactment of femicide laws was not merely a reaction to local conditions but part of a broader wave of transnational feminist activism that reshaped legal frameworks across Latin America. Thanks to this movement, 18 countries in the region have recognized femicide as a distinct hate crime (Sarmiento et al. 2014). In Mexico, feminist activists in political spheres, supported by international pressure from institutions such as the

---

[2] This estimate is based on the discrepancy between the number of likely femicides inferred from death certificates and the statistics reported by the judicial system, the Secretariado Ejecutivo del Sistema Nacional de Seguridad Pública (SESNSP).

[3] These certificates have also proven valuable in previous studies on femicide and violence (Olamendi 2017), due to their advantages and more likely accuracy over official estimates.

European Parliament, the UN, and the Inter-American Court of Human Rights, played a crucial role in condemning the Mexican government's impunity in handling these cases. These efforts ultimately resulted in the successful gradual enactment of femicide laws nationwide (Lagarde 2007; Lombard 2018). I demonstrate that, despite varying rates of violence, Mexican states exhibited similar pre-reform trajectories, supporting the parallel trends assumption necessary for a reliable difference-in-differences analysis.

This paper contributes to the literature on crime reduction interventions, particularly those targeting violence against women, by making five key contributions. First, it provides the first causal estimates of the impact of femicide laws on lethal violence against girls and women. Contrary to expectations that femicide laws would reduce lethal violence, the findings indicate no statistically significant impact on the number of femicides, female homicides, female suicides, or reported disappearances of girls and women. This conclusion holds in the short and long term and is robust to whether other laws were enacted alongside. The results also remain consistent when accounting for broader factors linked to gender-based violence, including firearm availability—proxied by the number of males killed by firearms—and household stressors, like state-level unemployment rates.

Second, this paper is the first to assess the impact of unilateral divorce laws while accounting for the presence of femicide legislation, addressing a key gap in understanding how legal frameworks on gender-based violence may interact. Mexico offers a unique setting due to the staggered adoption of both legal reforms, enabling a systematic assessment of their combined effects. Previous studies have focused solely on unilateral divorce laws, overlooking the potential role of femicide laws. For example, Hoehn-Velasco and Silverio-Murillo (2020) found that in Mexico unilateral divorce laws led divorce rates to increase by over 30% in the short run (2.5 years) without affecting homicides—a contrast with the homicide-reducing effects seen in Europe and the US (Brassiolo 2016; Stevenson and Wolfers 2006). García-Ramos (2021) also found that unilateral divorce laws led to a modest 3.7 percentage point rise in intimate partner violence in Mexico six to eight years later, especially among married women, suggesting divorce reforms could have affected homicides in the long run.

In line with these earlier studies, I find that the unilateral divorce reform led to an increase of divorce rates by over 30% soon after, and that such impact has remained for nearly eight years (90 months) post-intervention. Despite this persistent increase in divorces in the long run, there is no evidence that femicide laws—whether implemented alongside or independent of unilateral divorce laws—affected femicides, homicides of women, suicides, or disappearances of girls or women.

Extensive robustness checks also reveal no evidence that the introduction of equitable asset divorce compensation, which grants up to 50% of marital assets to the spouse responsible for household and childcare duties, has affected rates of lethal violence, suicides, or disappearances of women or men. While death certificates do not specify marital property regimes, the findings hold consistently across subgroups most likely to be affected, including those who were married, separated, or divorced, regardless of gender. These results add to the growing body of research on the effects of unilateral divorce laws on gender-based violence, suggesting that in contexts like Mexico, such legal reforms have neither systematically escalated nor mitigated lethal violence (Hoehn-Velasco and Penglase 2021b; García-Ramos 2021; Daby and Moseley 2023).

A third key contribution is the analysis of whether abortion decriminalization, when enacted alongside femicide legislation, impacts lethal gender-based violence. Prior research suggests that intimate partner violence can lead to non-consensual pregnancies, and paradoxically, that women seeking abortions may also face heightened risks of violence (Coyle et al. 2015; Pallitto et al. 2013; Muratori 2021; Aston and Bewley 2009). If these mechanisms hold, decriminalization of abortion could escalate lethal violence against women. To test this, I use the Mexican state of Oaxaca as a case study—an outlier where femicide laws preceded other legal reforms. The findings reveal no significant effect on femicides, female homicides, suicides, or disappearances relative to comparable states. These results reinforce the null impacts that femicide laws have. More importantly, these findings demonstrate that abortion decriminalization neither increases nor reduces lethal violence against women, offering critical insights for other settings considering similar policy reforms.[4]

The fourth key contribution is to investigate whether the impact of femicide laws, unilateral divorce, and abortion decriminalization differs across 17 subgroups of female homicide victims, defined by age, marital status, employment, education, pregnancy status, and Indigenous language use. This analysis offers key insights into the potential mechanisms by which these legal reforms may impact patterns of lethal violence by identifying which groups, if any, are likely to be most affected. Once again, the findings reveal no heterogeneity across these dimensions, reinforcing the conclusion that these legal reforms had no measurable impact on trends in lethal violence against women.

[4] These findings are particularly relevant as the Mexican Supreme Court unanimously ruled in September 2023 that penalizing abortion is unconstitutional.

Fifth, the paper contributes to the international literature on whether harsher prison sentences deter crime, including gender-based violence (Koppa and Messing 2021). I begin by demonstrating that, contrary to popular belief, not all Mexican states initially imposed harsher penalties for femicide than for homicide. [5] This evidence is crucial, as it highlights the risks for policymakers and activist groups in enacting femicide legislation without considering existing homicide penalties. This variation in sentencing across states and over time allows for comparative analysis, revealing that longer prison sentences for femicides do not significantly reduce femicides, female homicides, suicides, or even disappearances, which could be a means of concealing femicides. These findings are consistent over both short- and long-term periods and remain valid when comparing the severity of sentences for femicides and homicides across different states. This aligns with existing empirical literature, which shows that harsher penalties do not necessarily deter crime (Chalfin and McCrary 2017). But even more importantly, the findings indicate that even with stricter penalties, offender behavior does not change—there is no evidence of deterrence or strategic adaptation, such as an increase in disappearances to avoid prosecution.

The results hold across a range of additional robustness checks. Placebo tests using 24-month pre-treatment periods detect no significant pre-trends, reinforcing the credibility of the identification strategy. Alternative outcome measures, such as relative killing rates taking into account the population size of women instead of absolute casualty counts, produce consistent findings. More importantly, the results remain stable even when applying the synthetic difference-in-differences method proposed by Arkhangelsky et al. (2021) which mitigates potential biases by constructing a data-driven control group.

Overall, these findings are consequential to policymakers, as they show that femicide laws have no effect, at least in the context analyzed, where impunity is widespread. In Mexico, nearly 98% of all crimes, including homicides, remain unsolved (Human Rights Watch 2020), rendering legal reforms largely symbolic. A more comprehensive approach is needed—one that directly challenges impunity and the systemic injustices that allow gender-based violence to persist.

[5] By December 2020, all states had implemented harsher sentences for femicides than for homicides, except for five states where the penalties were equal. Nevertheless, there remained variation in both the average prison sentences for femicides and the differences in sentencing compared to homicides.

## 2. Background

*2.1 The surge of femicides and who are the killers?*

In the 1990s, Ciudad Juárez saw a sharp increase in gender-based killings, known as *las muertas de Ciudad Juárez*. The victims were often raped, mutilated, and left in public places, yet arrests were rare. Of the estimated 500 victims, only 20 perpetrators have been identified, most of whom were intimate partners or members of organized crime (Washington Valdez 2005). The reasons for Ciudad Juárez becoming a focal point for random killings of women are still debated. The city's role as a border manufacturing hub increased economic opportunities, but it also created a lawless zone where criminals operated with impunity (Livingston 2004).

During the late 1990s, security conditions began to deteriorate in several cities following the victory of opposition parties in local elections, which led to the unraveling of long-standing arrangements between drug cartels and law enforcement. Previously, under the Institutional Revolutionary Party (PRI), which ruled from 1929 to 2000, organized crime operated with relative stability, benefiting from tacit agreements that limited large-scale civilian violence. But the PRI's loss of the presidency in 2000, and again in 2006, both times to the National Action Party (PAN), led to escalating conflicts among criminal organizations. In response, President Felipe Calderón deployed 45,000 troops in a militarized campaign against drug cartels, aiming to dismantle their leadership. Although this 'war on drugs' strategy led to the capture or killing of 25 cartel leaders and 160 lieutenants, it also contributed to unprecedented levels of drug-related violence. Within six years, the national homicide rate doubled, rising from one of the lowest in Latin America to nearly 22 homicides per 100,000 inhabitants. While the conflict disproportionately affected men—who accounted for 90% of over 300,000 deaths thus far—it also fueled a sharp increase in female homicides, reversing prior gains in reducing violence against women.

Despite claims that some women have been targeted because have work ties to drug cartels, there is no strong evidence of such.[6] Instead, multiple studies have found a clear cause-and-effect relationship between the militarized strategy against drug trafficking organizations and the overall increase in homicides, unemployment rates, and criminality in the country (Dell 2015; Calderón et al. 2015; Gutiérrez-Romero 2016). The rise of overall homicides, including those of women, has also been noted in other armed conflicts, where financial hardship, deteriorating labor market opportunities, and heightened domestic tensions contribute to increased risks (Azariadis and Drazen 1990; Brück, Maio, and Miaari 2019). In Mexico, the increased availability of firearms, driven by criminals seeking to arm themselves for turf battles, has intensified lethal violence. Firearms are now responsible for 71% of all murders, up 20 percentage points from two decades earlier (Dube, Dube, and García-Ponce 2013).[7]

[6] More than 90% of crimes in Mexico go unreported, and from those reported, less than 4% reach a sentence. Thus, there are no reliable statistics available for these crimes to establish such links (Olevera Lezama 2023).

*2.2. Origin of femicide legislation*

Before the war on drugs, activists and lawmakers in Mexico had already expressed concerns about the increasing number of hate crimes against women no longer restricted to Ciudad Juárez. Through the efforts of transnational feminist activists, successful campaigns brought to light the severe negligence of the Mexican government (García-Del Moral and Neumann 2019). In response, the European Parliament issued three condemnations regarding the Mexican government's handling of these cases, while in 2005 and 2006 the UN and the Inter-American bodies urged Mexico to criminalize femicide. In 2009, the Inter-American Court of Human Rights ruled that Mexico had failed to prevent, investigate, and prosecute the abduction, brutal rape, and murder of three young female workers in Ciudad Juárez. This ruling confirmed that Mexico had violated multiple human rights and engaged in gender discrimination. In Mexico, legislator Marcela Lagarde led a congressional commission to examine the roots of gender-based violence, concluding that these crimes were not isolated incidents, but deliberate murders driven by gender-based hatred. Drawing from Diana Russell's (1977) concept of femicide, Lagarde translated it as 'feminicidio' to emphasize not only the killings but also the state's complicity in failing to prevent and prosecute them (Lagarde 2007).[8]

In Mexico, states can either adopt federal criminal code laws or establish their own within state-level criminal codes. In 2010, Guerrero became the first state to explicitly criminalize feminicide, requiring distinct investigative and prosecutorial procedures.[9] Then, in 2012, the federal criminal code defined feminicides as hate killings of women because of their gender, regardless of age. By 2019, all 32 states had incorporated femicide legislation into their criminal codes, with most adopting similar criteria for classification but varying widely in prison sentencing severity. Despite the wide use of the term feminicide in the law, and in popular discourse, holding the government accountable for addressing gender-based violence has been nearly impossible to prosecute in practice. Thus, for the rest of this article, this term will be translated into English as femicide.

[7] In response, the Mexican government has filed a lawsuit against US gun manufacturers, accusing them of facilitating illegal arms sales to traffickers.

[8] As a result of these efforts, the General Law for Women's Access to a Life Free from Violence was enacted in 2007, introducing the term 'feminicide violence' for the first time. This law defines this type of violence as 'misogynistic behaviors that can lead to social and legal impunity, ultimately resulting in homicide and other forms of female death' (Secretaria de Servicios Parlamentarios, 2007, p.13).

## 3. Data

### *3.1. Data on femicide, homicide, divorce, and abortion decriminalization legislation*

A comprehensive dataset was compiled to track the implementation dates of femicide laws across the 32 states' criminal codes. According to federal law, femicide is defined as the killing of a girl or woman due to her gender, provided that at least one of the seven specific conditions listed below is met. The criteria are not ranked in any particular order, but each serves to distinguish femicide from other homicides. Broadly speaking, in 31 of Mexico's 32 states, the definition of femicide has evolved to align with the definition provided by federal criminal law. The only state with significant differences is Michoacán, which does not consider kidnapping as an element to classify a killing as femicide.

- Signs of sexual violence.
- Non-lethal humiliating injuries, such as mutilations, before and or after death, including acts of necrophilia.
- Harassment or threats before being killed.
- Domestic violence or violence from a family member, coworker, or schoolmate.
- A close, familial, or intimate relationship with the perpetrator.
- The body is left in a public place, such as a highway, park, or deserted area.
- Isolation or inability to communicate due to being kidnapped before death.

[9] Femicides are prosecuted ex officio, requiring immediate investigation upon incident report. The Public Ministry coordinates the investigation and identifies those responsible. Unlike first-degree murder, femicide investigations must consider factors making women vulnerable and adhere to state guidelines (Olevera Lezama 2023).

*Date of Early Legal Recognition of Gender-Based Homicide.* Before the adoption of femicide laws, six states introduced harsher penalties for homicides of women motivated by misogyny and gender-based hate. In evaluating the impact of femicide laws, I treat the implementation date of these penalties for gender-motivated homicides as the effective date of femicide laws in these six states. I do so as this date signifies an early legal recognition of gender-based discrimination in homicide cases. Chihuahua, home to Ciudad Juárez and the focal point of the femicide debate, was the first to adopt such measures in 2003. However, despite leading in sentencing reforms, it was among the last states to formally incorporate femicide into its criminal code, doing so only in October 2017 (Table A.1).

*Penalties on Femicides and Homicides Laws.* —The dataset also records the minimum and maximum sentences for both femicides and homicides from the criminal codes of all 32 states, tracking changes over time. Under the Mexican criminal code, homicides are broadly classified into intentional and unintentional. Notably, when femicide cannot be proven, homicide penalties apply, as explicitly stated in several state criminal codes. Thus, in the robustness check section, I compare femicide penalties with those for first-degree intentional homicides (referred to as 'calificados' in Spanish), which involve aggravating factors such as advantage or treachery. This comparison provides a good benchmark for assessing the penalties murderers would face in the absence of femicide laws.

Criminal codes define minimum and maximum prison sentences for crimes. To estimate the average prison sentence for femicide each month, I take the mean of the minimum and maximum sentences for each state, applying the same method to homicides. Contrary to common belief, average femicide sentences have not always exceeded those for homicides; in Baja California, Chihuahua, Coahuila, and Tlaxcala, homicide sentences were initially longer. These penalties, however, have evolved. By December 2020, as shown in Figure 1, all states imposed longer average sentences for femicide, except for Michoacán, Quintana Roo, Estado de México, and Sinaloa, where the average sentences for both crimes remained identical.[10]

[10] In five states in the country, Estado de México, Nayarit, Durango, Oaxaca, and Sinaloa, even higher maximum prison sentences are imposed for femicides if aggravating factors are present. These factors include situations where the perpetrator had a trusted relationship with the victim. When these aggravating higher sentences are considered, the average penalty for femicides in

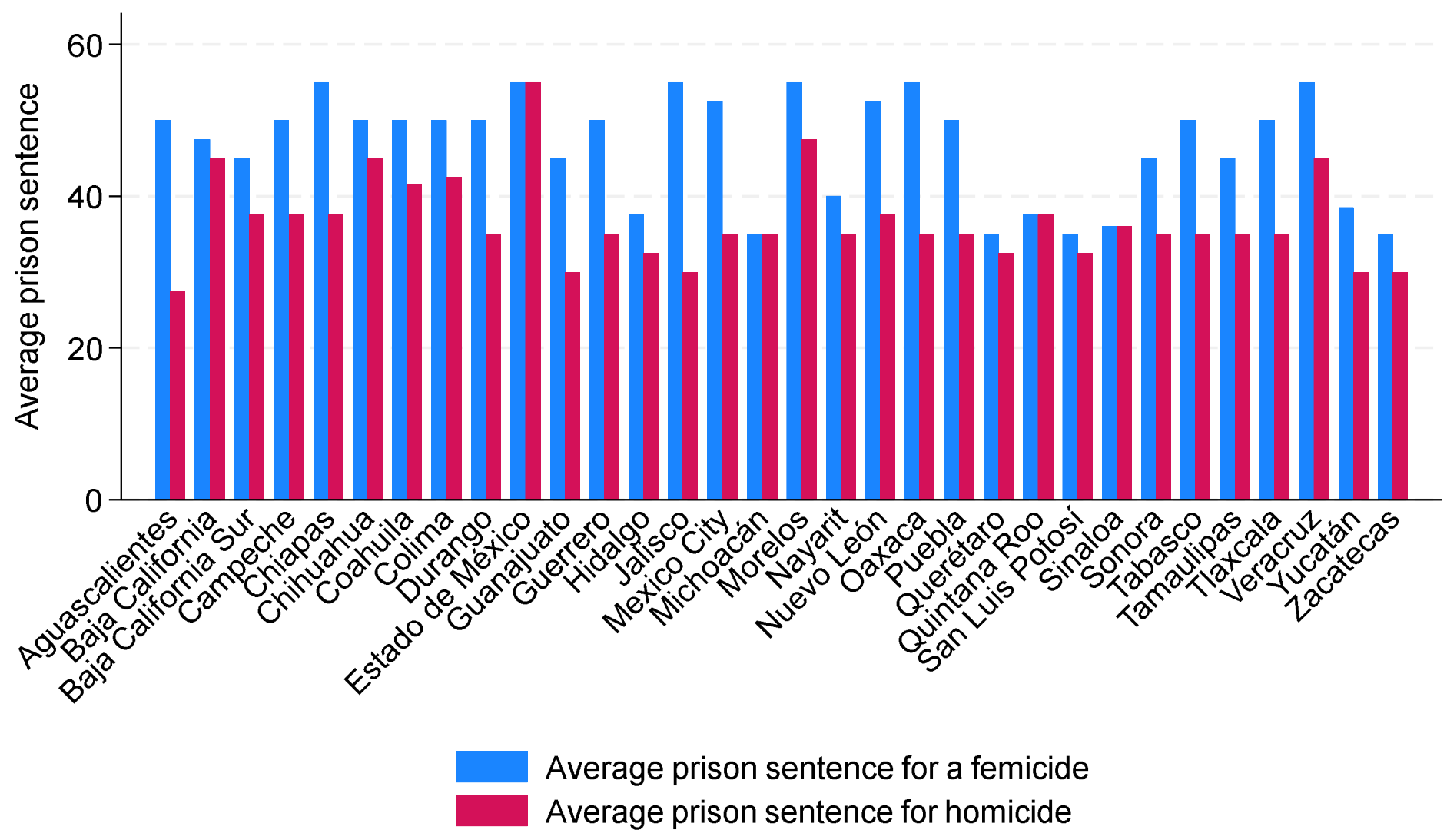

**Fig. 1.** Average prison sentences for femicides and homicides for the year 2020

*Unilateral Divorce, Asset Compensation, and Decriminalization of Abortion Laws.*— During the period of analysis, 2000-2020, twenty-six states gradually enacted unilateral divorce laws, while twenty-three introduced provisions requiring that, upon divorce, 50% of marital assets be transferred to the spouse primarily responsible for household and childcare duties when the couple had originally married under a separate property regime. Additionally, two states decriminalized abortion. Table A.1 details the dates when each state enacted femicide laws (or laws enhancing penalties for gender-based homicides), decriminalized abortion, or introduced unilateral divorce laws.

As shown in Table A.1, Mexico City stands out as the first authority to enact all four legal reforms. It introduced new asset compensation in divorce settlements in 2000, decriminalized early-term elective abortion in 2007, implemented unilateral divorce in 2008, and passed femicide laws in 2011. The provision of free access to elective abortions in public hospitals and clinics resulted in a decline in maternal morbidity and reduced fertility rates in the city (Gutiérrez Vázquez and Parrado 2016; Clarke and Mühlrad 2021). Despite these promising outcomes, this progressive shift did not trigger widespread policy adoption but instead provoked a backlash. Between 2000 and 2020, only Oaxaca followed in decriminalizing abortion—seven years after implementing femicide laws and nearly two years

---

Sinaloa and Estado de México would be higher compared to that for homicides. However, it is important to note that this increased penalty does not apply to all types of femicides.

after introducing unilateral divorce. Given this sequencing, Oaxaca is analyzed separately to assess the combined impact of gradually implementing these laws following the enactment of femicide laws.

### *3.2. Data on femicides, homicides, suicides, disappearances and divorce rates*

Mexico's official record of femicides began in 2015, as reported by the Secretariado Ejecutivo del Sistema Nacional de Seguridad Pública (SESNSP). This official record-keeping started five years after the femicide law was enacted in Guerrero and after such laws had already been adopted by nearly all states. Due to the lack of official femicide data before the enactment of these laws and issues already mentioned with data reliability, I use four alternative proxies instead.

The first three proxies refer to the number of presumed homicides of women (regardless of the perpetrator's motive or intent, but excluding accidental deaths), likely femicides and female suicides, estimated directly from official death certificates and aggregated monthly at the municipal level from 2000 to 2020. Death certificates are issued by authorized forensic medical services before any investigation is carried out by public ministries, making this information less vulnerable to the inefficiencies of the judicial system.[11] Families of the deceased are under significant pressure to obtain these certificates, as they are crucial documents required for various processes, including carrying out burials, dealing with inheritances and life insurance (Olamendi 2017). The anonymized death certificates are publicly released by the National Institute of Statistics and Geography (INEGI). Fetal deaths are excluded, as they are not recorded in the death certificates analyzed.

Official death certificates do not directly differentiate whether a homicide can be categorized as a femicide. However, I distinguish the likely femicides from presumed homicides by filtering the rich information contained in these certificates using the criteria set by the femicide law. I identify femicides as those homicide victims who experienced at least one of the following four conditions: sexual assault, a kinship/trust relationship with the alleged aggressor, domestic or non-domestic violence as directly classified in the certificate, or if the body was found in a public place. These conditions are considered in both the federal criminal code and by all 32 states.

[11] Authorized forensic medical services issue these death certificates when a death occurs in a hospital or clinic. By law, for any other death caused by an accident, violence, or likely homicide, the relevant ministry agency must be informed so that the competent authority can provide the corresponding death certificate and launch any necessary investigation.

Since death certificates are completed before any investigation takes place, only 1.4% of all certificates from the years 2000 to 2020 across the entire country include information about the alleged relationship with the likely assassin. All these 1.4% cases would still be classified as femicides based on the other criteria used. While death certificates serve as a useful alternative measure of femicides, it is important to acknowledge their limitations. For example, only 8.8% of victims are classified as having experienced domestic violence, and an additional 5.6% are classified as having experienced non-domestic violence. These low statistics may be because death certificates are provided by medical physicians or authorized health authorities, rather than through investigations conducted by prosecutors. Partners or relatives are also asked to provide additional information to complete the death certificate of their relative. However, they may be reluctant to disclose details related to domestic violence particularly if they are directly involved in these incidents. Another limitation is that death certificates do not include information about three critical characteristics relevant to determining the likelihood of femicide, as outlined in federal and state criminal codes. These characteristics are whether the victim experienced threats, degrading wounds, or was kidnapped.[12] While some of these cases may already be captured by the proxy indicator of femicides built based on other characteristics of the victim, the reports of disappeared people—analyzed separately, as explained next—may also reveal additional femicide cases.

Due to the potential incentives femicide legislation creates for concealing women's murders, the number of disappeared women may have been affected. As a fourth indicator, I analyze the number of reported disappearances of girls, teenagers, and women likely to be crime victims. These data come from the Registro Nacional de Personas Desaparecidas y No Localizadas (RNPDNO), which records official reports of disappearances. Such reports can be filed by family members, friends, or anyone with knowledge of the case.

I also evaluate the impact of unilateral divorce laws on divorce rates. To do so, I use INEGI's individual divorce statistics at the municipal and monthly levels during 2000-2020, which indicate whether the divorce was conducted via administrative or judicial proceedings. I calculate the divorce rate as the number of divorces per 1,000 people. The total population, also at the monthly and municipality levels, used for this rate is obtained from Consejo Nacional de Población (CONAPO).

[12] As mentioned earlier, Michoacán is notably the only state that does not consider kidnapping as an aggravating factor.

### *3.3. Data on controls*

In the main analysis, Section 5, I use three key control variables. That is the monthly number of male homicides by firearms at municipality level. This indicator is obtained from death certificates provided by INEGI. I also include the state-level unemployment rates for women and men at quarterly levels. These unemployment indicators are sourced from the National Occupation and Employment Survey (ENOE), the country’s largest labor survey conducted quarterly by INEGI. In the robustness section, I test alternative controls, excluding unemployment rates, and instead incorporating satellite night light intensity at the monthly and municipality level. The data are sourced from the Earth Observation Group, Payne Institute for Public Policy  (Hansen et al. 2013).

### *3.4. Data overview*

Homicides of women have consistently accounted for less than 20% of total homicides. Yet, their absolute number—and the rise of femicides—has grown steadily since 2006, coinciding with the militarization of Mexico’s fight against drug trafficking organizations. Figure 2 illustrates this pattern, comparing official femicide records with my estimated femicide proxy based on death certificates. While the official classification of femicides only began in 2015, the proxy suggests a significantly higher incidence, rising in tandem with overall female homicides.

Figure 2 also shows the rise in the number of reported disappearances of women, who make up 30% of the more than 100,000 cases that have been reported since 2000, with most cases being Mexican nationals rather than international migrants. Disappearances have increased rapidly since 2018. The reasons for this shift remain unclear, though changes in criminal strategies appear likely. According to Crisis Group (2024), authorities and criminal organizations have struck informal pacts, encouraging criminal groups to reduce overt violence to ease public pressure. Such arrangements may have incentivized a transition from homicides to disappearances.

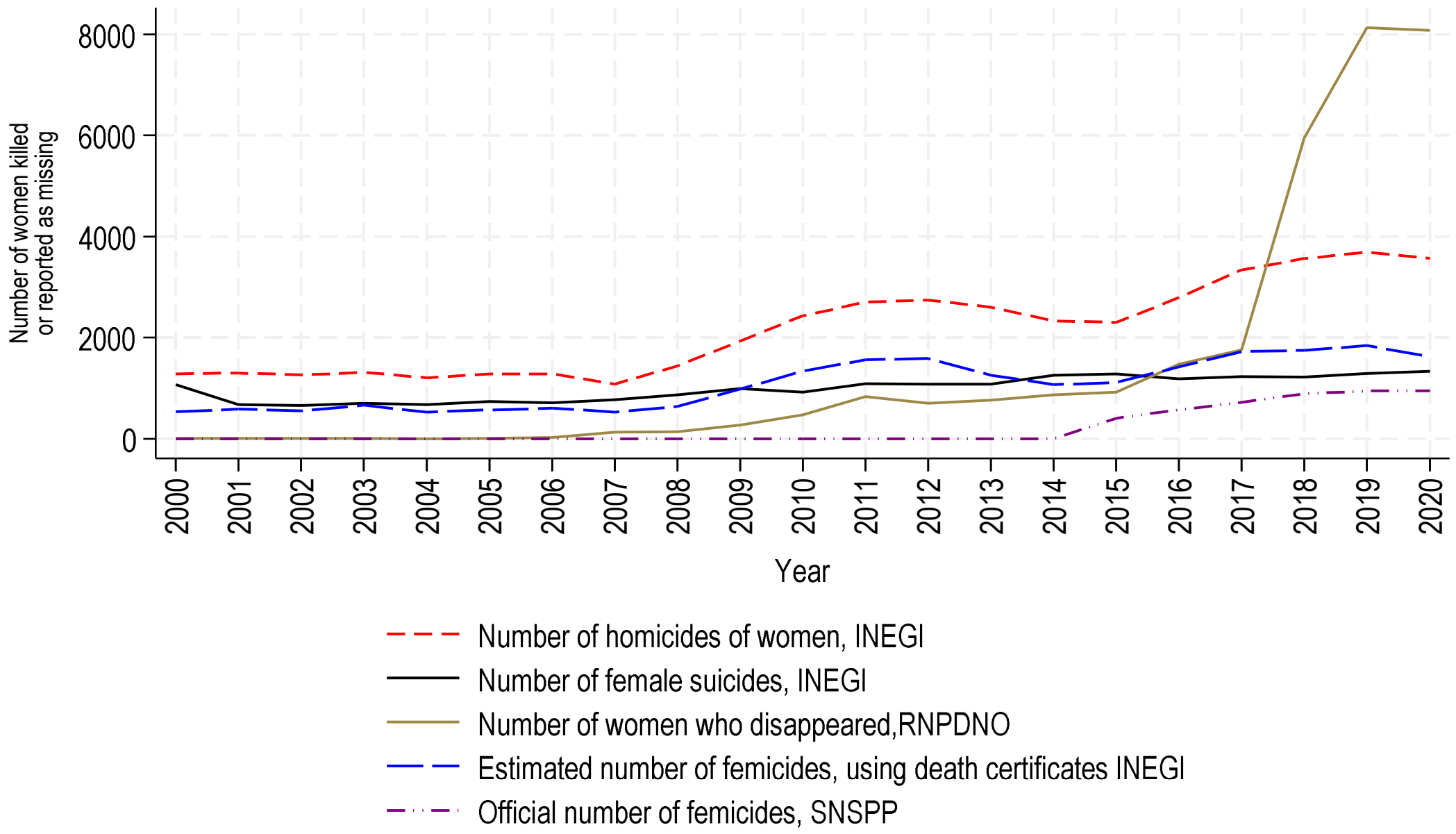

**Fig. 2**. Number of homicides of women, femicides, and number of disappearances of women
*Notes:* The number of homicides, femicides, and suicides are calculated using death certificates available from INEGI. The number of disappearances is sourced from RNPDNO.

Appendix Figure A.1 presents instead the homicide, suicide, and disappearance rates per 100,000 inhabitants, disaggregated by gender. As previously noted, the 2006 launch of the war on drugs marked a turning point, doubling in just six years the overall homicide rate from nearly 10 to nearly 22 murders per 100,000 inhabitants. This surge was not a temporary fluctuation but the start of a prolonged security crisis that continues to define Mexico's landscape of violence.

Figure 3 shows the geographic distribution of the average rate of homicides, disappearances, and femicides during 2000-2020, revealing both the scale of the crisis and its uneven spread. Female homicide rates tend to be highest in regions with high rate of male homicides. Similarly, disappearances—of both men and women—tend to concentrate in the same areas. While northern cities have long been synonymous with cartel-related violence, femicide hotspots and disappearance clusters have steadily multiplied across the country. Figure A.2 shows this trend, tracking quinquennial femicide rates and showing how new centers of gender-based violence have continued to evolve.

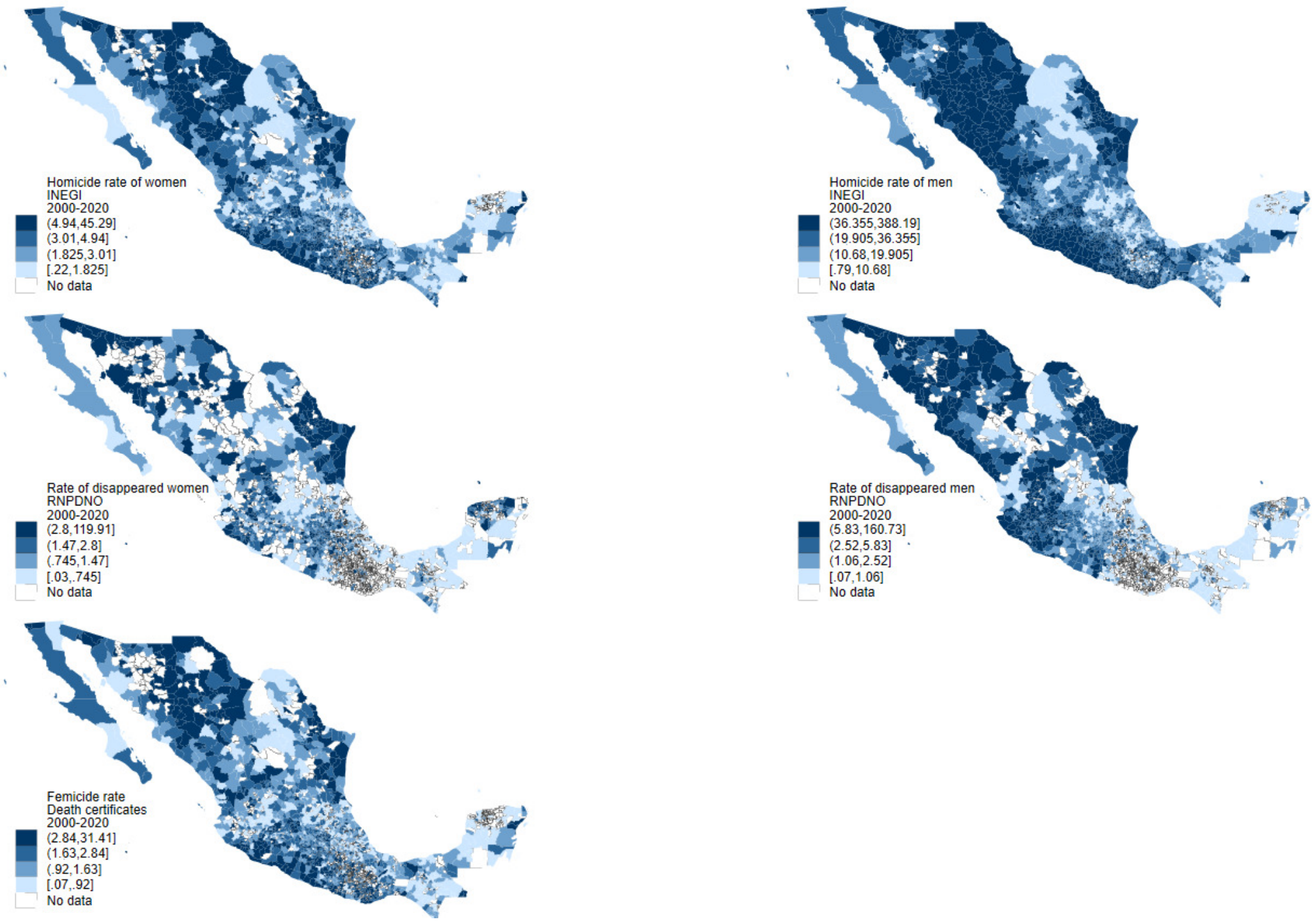

**Fig. 3**. Geography of average rate of homicides, femicides, and disappeared people during 2000-2020

In the rest of the paper, I concentrate on absolute numbers rather than rates because such high levels of male and female homicides can have a substantial impact on population size, especially in smaller localities. This approach allows for a more direct assessment of femicides, homicides of women, suicides, and reported disappearances without distortions caused by varying population sizes. Nonetheless, in the robustness section, I show that results remain robust if using killing rates instead.

## 4. Empirical Strategy

I use the staggered difference-in-difference (DID) estimator, proposed by de Chaisemartin and D'Haultfoeuille (2022) to evaluate the impact of each relevant legislative reform. As shown in equation (1) this estimator, denoted by $DID_{g,l}$ compares the evolution of an outcome variable, $Y$, at time periods, $t$, between groups, $g$, that have been treated, $D$, and those who have not yet been treated from period $F_g$-1 to $F_g+l$. compensation.

I begin by analyzing the impact of unilateral divorce laws—commonly known as *no-fault divorce*—on divorce rates, which are measured monthly at the municipality level. However, the primary outcomes of interest, $Y$, are the number of femicides, homicides of women, female disappearances, and female suicides, each measured monthly at the municipal level. For these outcomes, the treatment *D* refers to whether municipalities are in states that have enacted femicide laws.

I evaluate both the short-term and long-term effects of femicide laws, distinguishing between cases where they were enacted independently and those where they coincided with other policy changes. I address potential confounding effects by separately examining states that enacted exclusively femicide laws without any other reform. Then I focus on those states that enacted femicide laws along with unilateral divorce laws, decriminalized abortion, or the new asset divorce reform.

The analysis includes standard time and area fixed effects, as well as various control sets detailed in the primary analysis and robustness checks. To estimate $\text{DID}_{g,l}$ the standard errors are clustered and bootstrapped at the state level.

$$DID_{g,l} = Y_{g,F_{g+l}} - Y_{g,F_{g-1}} - \sum_{g':D_{g',1=0},F_{g'}>F_{g+l}} \frac{N_{g',F_{g+l}}}{N^u_{F_{g+l}}} \left(Y_{g',F_{g+l}} - Y_{g',F_{g-1}}\right) \quad \text{Eq (1)}$$

where $N_t^u = \sum_{g:D_{g,1=0},F_g>t} N_{g,t}$ represents the number of observations in the untreated groups at period *t*, starting from period 1 up to t.

One of the advantages of the $\text{DID}_{g,l}$ estimator is its robustness to heterogeneous treatment effects across groups and over time. This estimator is based on the assumption that, in the absence of treatment, both the treatment and comparison groups would have experienced the same outcome evolution. This assumption can be indirectly tested by assessing whether the treated and untreated groups had parallel trends in outcomes before implementing the treatment. To this end, placebo estimators are estimated. These compare the outcome evolution between the treated and untreated groups before group *g* receives treatment, from period $F_g$- *l*.-2 to period $F_g$-*1*. These 'long-difference' placebos make it easier to detect if the compared groups had different trends over multiple periods rather than just in two consecutive periods.

## 5. Results

### *5.1 Impact of unilateral law on divorces*

I start by estimating the impact of the unilateral divorce laws on divorce rates in Mexico, using the staggered DID estimator. The treatment group comprises municipalities within states that enacted unilateral divorce laws, while the control group comprises municipalities in states that have yet to adopt such legislation. The outcome variable is the natural logarithm of the judicial divorce rate.

In Mexico, divorces can occur via administrative or judicial proceedings. Administrative divorces are reserved for cases where both spouses mutually agree to end their marriage. Judicial divorces, by contrast, occur when one spouse seeks to terminate the marriage unilaterally (in treatment states) or must cite specific legal grounds for divorce (as in control states).

Table 1 presents the results, which show a statistically significant increase in judicial divorce rates attributable to unilateral divorce laws. I use four alternative specifications. The first one has no controls other than area and fixed effects. The second one adds the number of males killed by firearms, as a proxy for firearm accessibility and local conflict that may affect domestic violence and the overall risk of femicides (Nguyen and Le 2022; La Mattina 2017; Stojetz and Brück 2023). This second specification also adds the state-level unemployment rate for both females and males. I use these indicators based on earlier literature showing that domestic violence is affected by unemployment, and, more importantly, that unilateral divorce laws in Mexico have not impacted women's labor market participation (Meyer et al. 2024; Bobonis, González-Brenes, and Castro 2013; Hoehn-Velasco and Penglase 2021b; 2021a). To account for the possibility that rising divorce rates may, over time, impact unemployment, the third specification lags all controls by 12 months. The fourth specification lags all controls by 24 months.

**Table 1**

The difference-in-differences estimate on the judicial divorce rate

| | (1) | (2) | (3) | (4) |
|---|---|---|---|---|
| DID estimate | 0.302*** | 0.302*** | 0.301*** | 0.303*** |
| | (0.068) | (0.068) | (0.068) | (0.068) |
| Number of observations x periods | 239065 | 239065 | 239065 | 239065 |
| Number of switchers x periods | 18557 | 18557 | 18557 | 18557 |
| Number of months used for post-treatment | 90 | 90 | 90 | 90 |
| Controls | | | | |
| Area and time fixed effects | Yes | Yes | Yes | Yes |
| Unemployment rates of males and females, the number of males killed by firearms | | Yes | | |
| Unemployment rates of males and females, the number of males killed by firearms, all lagged by 12 months | | | Yes | |
| Unemployment rates of males and females, the number of males killed by firearms, all lagged by 24 months | | | | Yes |

*Notes*: Standard errors, clustered at the state level and bootstrapped, in parentheses. Significance levels *** p<0.01, ** p<0.05, * p<0.1.

Earlier studies on Mexico have shown that unilateral divorce laws led to a short-term increase in divorce rates within the first 2.5 years of implementation, with effects persisting for up to eight years (García-Ramos 2021; Hoehn-Velasco and Penglase 2021b). Consistent with these findings, using the staggered DID, I observe a significant rise in judicial divorces soon after the reform and even 90-month post-enactment period. Across all specifications, unilateral divorce laws are associated with an approximately 35% increase in judicial divorces. Figure 4 illustrates the dynamic effects of the second specification, which follows a trajectory closely aligned with the other specifications. This event study illustrates standard errors with red vertical bars around the estimates, showing no evidence of a violation of parallel trends before treatment.

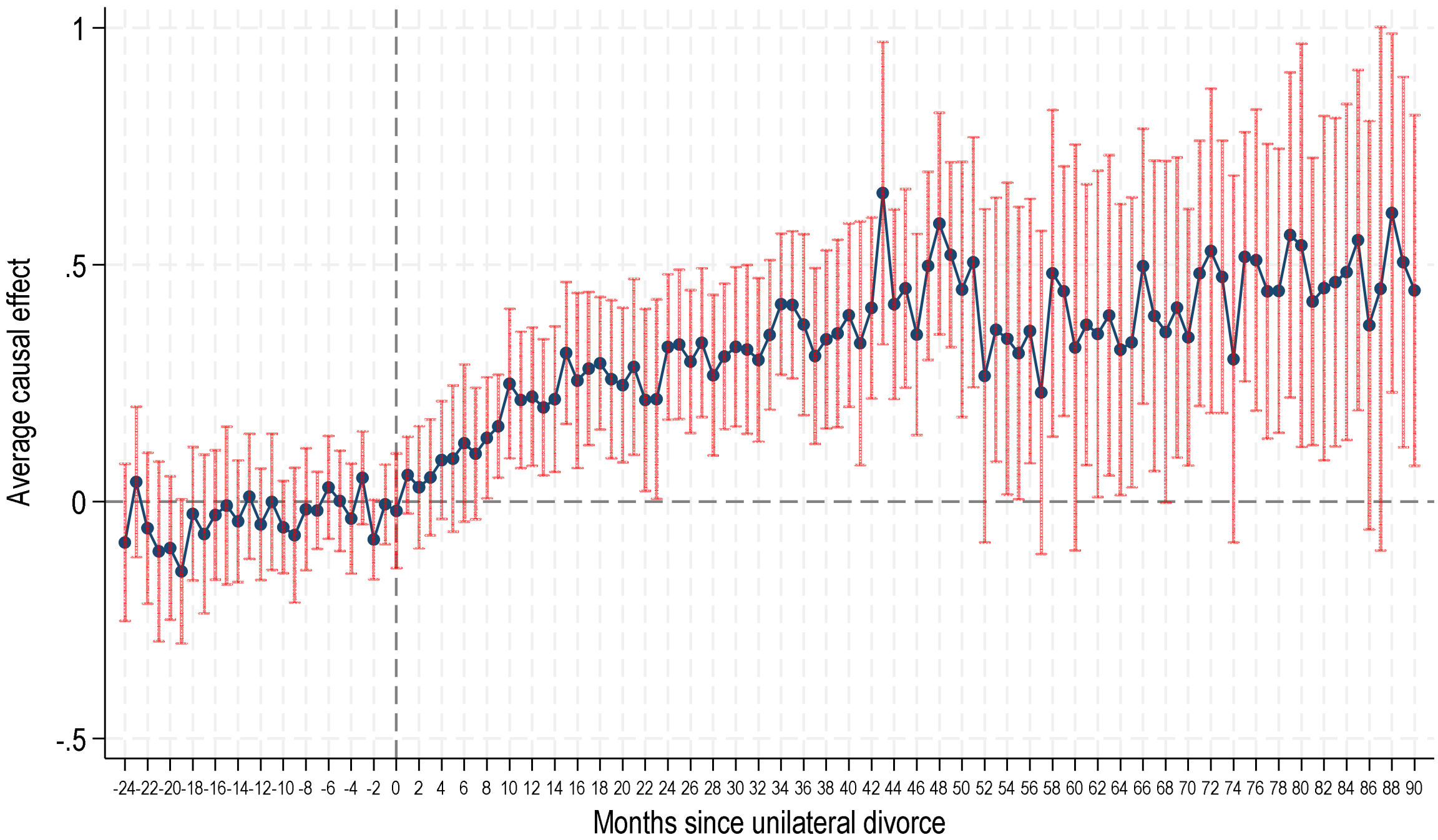

**Fig. 4.** The difference-in-difference on the divorce rate, column 2 of Table 1

These findings suggest a transformative effect of unilateral divorce laws in reducing barriers to marital dissolution. The results align with theoretical predictions from family economics, which suggest that lowering the costs of marital exit increases the likelihood of divorce, at least in the short-run (Wolfers 2006). However, the broader implications for women's safety are less clear.

As mentioned earlier, previous research has found no significant impact of unilateral divorce laws in Mexico on female homicides (Hoehn-Velasco and Silverio-Murillo 2020). However, this earlier work focused on only 2.5 years post-unilateral divorce reform and overlooked the fact that, concurrently, all states implemented femicide laws, most also had new equitable asset compensation upon divorce, and a minority decriminalized abortion. These legal changes may have also influenced trends in female homicides and other types of lethal violence, thereby obscuring the distinct effects of each law, an aspect I analyze next.

*5.2 Impact of femicide laws on femicides*

In this sub-section, I use the staggered DID estimator to assess the impact of femicide laws on femicide counts. The analysis includes area and time fixed effects as controls along with the unemployment rates of women and men (at state and quarterly levels) and the number of males killed by firearms (at municipal and monthly levels). I add these controls based on prior research which suggests economic downturns can heighten financial stress, potentially influencing intimate partner violence (Bhalotra et al. 2021; Anderberg et al. 2016). Additionally, firearm-related male deaths signal broader societal violence that can spill over, increasing the risk of generalized violence for women (Azariadis and Drazen 1990; Brück, Maio, and Miaari 2019). In the robustness section I show the results are stable even when using alternative control specifications.

In Table 2, column 1, I begin by isolating the effect of femicide laws on femicide counts, comparing municipalities in states that enacted these laws to those in states that have not yet adopted them. To prevent confounding, states that also implemented unilateral divorce laws or decriminalized abortion are excluded, leaving a sample of 220 municipalities across six states: Baja California, Campeche, Chihuahua, Guanajuato, Sonora, and Tabasco. Although this comparison is stringent—given that most states in Mexico have implemented unilateral divorce laws at some point, thus excluded—it remains the most transparent. Moreover, collectively, these states represent a substantial population of over 20 million inhabitants.

The estimated long-term DID effect on femicide counts is statistically insignificant over the 90-month post-implementation period, Table 2, column 1. Placebo tests, conducted for the 24 months preceding the introduction of femicide laws, also show no significant differences, supporting the validity of the DID approach (Figure 5).

**Table 2**

The difference-in-differences on the number of femicides

| | Impact of femicide law | Impact of unilateral divorce in states with femicide law | Impact of decriminalizing abortion in states with femicide, unilateral divorce and equitable divorce asset laws |
|---|---|---|---|
| | (1) | (2) | (3) |
| DID estimate | 0.037 | -0.013 | 0.038 |
| | (0.044) | (0.019) | (1.294) |
| Number of observations x periods | 28582 | 552505 | 9585 |
| Number of switchers x periods | 8117 | 58866 | 8550 |
| Number of months used for post-treatment | 90 | 90 | 14 |
| Controls: unemployment rates of males and females, the number of males killed by firearms, area and fixed effects | Yes | Yes | Yes |
| Treatment | Enacting femicide law, without unilateral divorce and abortion decriminalization | Enacting unilateral divorce in addition to femicide laws | Decriminalizing abortion, after enacting femicide, unilateral divorce laws and equitable divorce asset compensation |
| Comparison | Without femicide laws, unilateral divorce, or abortion decriminalization | With femicide law, but without unilateral divorce or abortion decriminalization yet | With femicide and unilateral divorce laws, each implemented around the same time |

*Notes*: Standard errors, clustered at the state level and bootstrapped, in parentheses. Significance levels *** p<0.01, ** p<0.05, * p<0.1.

To visualize the impact of femicide legislation, dynamic treatment effects are estimated for each of the 90 months analyzed, including extended placebo tests. The reference period for event-study regressions is set one period before treatment, following standard practice. Figure 5, Panel A, shows post-reform, femicide counts do not decline over the 90-month period. If anything, a small increase appears around period 85 but quickly loses statistical significance.

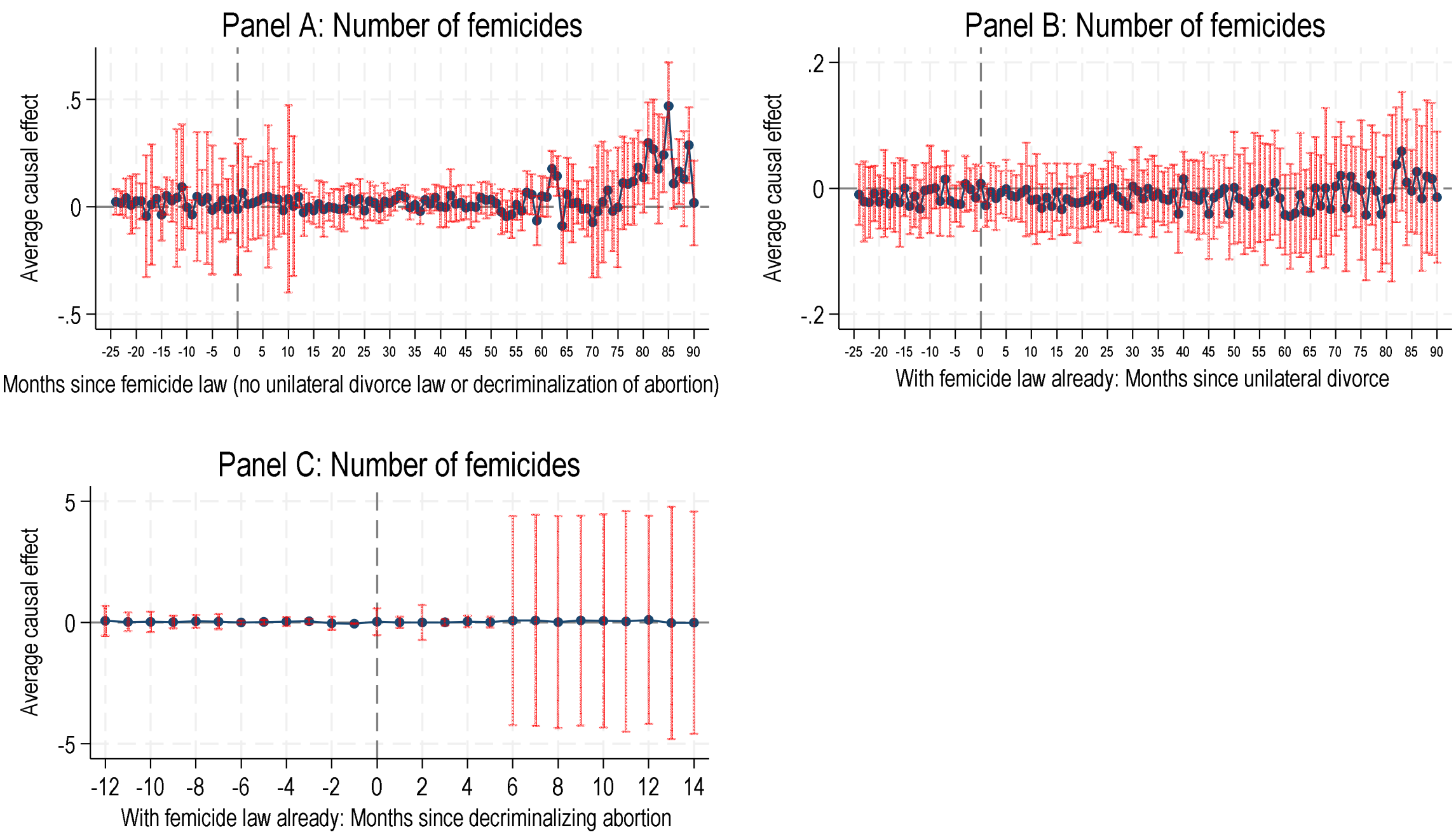

**Fig. 5.** The difference-in-difference on the number of femicides

Then, I conduct an even more stringent test by refining the sample of states. Among the six states analyzed so far, I exclude Baja California and Guanajuato, as both implemented equitable divorce asset compensation years before enacting femicide laws. I also exclude Chihuahua, which introduced aggravated penalties for gender-motivated homicides as early as 2003 but did not formally classify femicide as a separate offense until late 2017. This raises ambiguity about when potential perpetrators perceived an increase in penalties—whether in 2003, as assumed in earlier tests, or only after femicide was codified. So, I restrict the analysis to Campeche, Sonora, and Tabasco (with a combined population of 6.39 million people) the only three states in the country that enacted femicide laws without ever introducing unilateral divorce laws, equitable divorce asset compensation, or abortion reforms. This analysis is limited to a span of 15 months, which is the time between the implementation of femicide laws in the first state, Tabasco, and the last state, Sonora, from early 2012 to late 2013.

As shown in Table A.2, column 1, this narrower test yields a small but statistically significant increase in femicides (0.035). The dynamic effects are shown as well on Figure A.3, Panel A. This means that even under this highly restrictive specification, there is no evidence that femicide laws reduced femicides. Moreover, as discussed further below, there is no evidence in this specification that femicide laws had any statistically significant impact on female homicides, disappearances, or suicides (Table A.2, columns 2-4).

Then I move to examine whether the effectiveness of the femicide laws depends on the presence of unilateral divorce laws. This test is crucial as, if it is the case that femicide laws reduce intimate partner homicides only when paired with easier access to divorce, it would suggest that legal exit options play a critical role in protecting at-risk individuals. In this new test, shown in Table 2, column 2, the treatment group consists of municipalities in states that introduced unilateral divorce laws after femicide laws, while the comparison group includes municipalities where femicide laws were enacted but unilateral divorce laws had not yet (or never) been implemented.[13] The sample includes 1,454 municipalities across 22 states, tracked monthly.[14] These states collectively account for over 92 million residents, representing 73% of Mexico's population.

[13] In July 2015, Mexico's Supreme Court ruled that requiring proof of fault for divorce without mutual consent is unconstitutional, effectively mandating unilateral divorce nationwide. However, six states have not yet reformed their family laws, so fault-based divorces remain in their family code and in practice in these states.

Note that in this test, the DID estimator captures the joint effect of femicide and unilateral divorce laws rather than their independent impacts, meaning one may either amplify or offset the other. Table 2, column 2, shows that the combined implementation of femicide and unilateral divorce laws has no statistically significant effect on femicide counts over the 90 months following the introduction of unilateral divorce laws in states that had already enacted femicide laws.

Panel B of Figure 5 further confirms that the parallel trends assumption holds, with no significant differences in pre-intervention trends across any of the 24 pre-treatment periods. Moreover, the DID estimator remains statistically insignificant across all 90 post-intervention periods, reinforcing the robustness of the null result.

Next, I turn to another key question: Could the decriminalization of abortion, when introduced alongside equitable divorce asset compensation, unilateral divorce, and femicide laws, reduce femicides? To explore this, I analyze the case of Oaxaca, where femicide was implemented in 2012, unilateral divorce law and equitable asset compensation both in 2017, and abortion later decriminalized in 2019. I compare municipalities in Oaxaca to those in Quintana Roo and San Luis Potosí—the only two states that enacted femicide and unilateral divorce laws and equitable divorce asset compensation in the same sequence and timeframe but had not yet decriminalized abortion.[15] Oaxaca has a population of approximately four

---

[14] The states analyzed are Baja California, Campeche, Chiapas, Chihuahua, Coahuila, Colima, Estado de México, Guanajuato, Guerrero, Jalisco, Morelos, Puebla, Querétaro, Quintana Roo, San Luis Potosí, Sinaloa, Sonora, Tabasco, Tamaulipas, Tlaxcala, Veracruz, and Zacatecas. I excluded Mexico City and Oaxaca as they decriminalized abortion during the analysis period, as well as states that implemented unilateral divorce laws before femicide laws (Aguascalientes, Baja California Sur, Durango, Hidalgo, Michoacán, Nayarit, Nuevo León, and Yucatán). From all 22 states considered in this test, 16 had also enacted equitable divorce asset compensation.

[15] I exclude Mexico City, which decriminalized abortion in 2007 before implementing unilateral divorce laws in 2008 and femicide laws in 2011. Puebla is also excluded, despite implementing femicide and unilateral divorce laws around the same time as Oaxaca, due to its proximity to Mexico City, which may have provided earlier access to abortion.

million, while Quintana Roo and San Luis Potosí together have fewer than five million inhabitants.

I examine only the 14 months following abortion decriminalization due to its relatively late implementation in 2019. I conduct placebo tests using data from only the 12 months prior the abortion decriminalization because the DID estimation package restricts placebo periods to no longer than the post-treatment period minus two periods.

As shown in Table 2, column 3, decriminalizing abortion had no statistically significant effect on femicide counts. Figure 5, Panel C, further confirms that femicide numbers remained unchanged in the 14 months after decriminalization. Additionally, there is no evidence of parallel trends violations in the pre-intervention months, as the standard error bars are quite tightly centered around the estimate of zero differences in pre-trends.

*5.3 Impact on homicides of women*

Using death certificates to measure femicides provides a credible proxy for gender-based killings, as they rely on forensic medical evidence. However, this approach may underestimate the total number of femicides due to the absence of further judicial investigation beyond the issuance of these certificates. To address this limitation, I next examine the impact of femicide laws on the overall number of homicides of women, regardless of whether they present characteristics of a femicide or not.

Firstly, I estimate the DID impact of femicide laws by comparing changes in the number of homicides of women between states that enacted femicide laws—without concurrent unilateral divorce laws or abortion decriminalization—to similar states that had not yet adopted femicide laws. I use the same set of controls as in the analysis of the number of femicides. As shown in Table 3, column 1, the introduction of femicide laws had no statistically significant effect on the number of homicides of women during the 90 months post-policy reform. This result holds even under a more restrictive test that excludes states implementing equitable divorce asset compensation (see Table A.2, column 2, and Figure A.3, Panel B).

I next analyze whether the introduction of unilateral divorce laws in states that had already implemented femicide laws affected the number of female homicides. The DID estimate, presented in Table 3, column 2, indicates no statistically significant effect during the 90 months post-policy reform.

Further, I examine separately, in Table 3, column 3, the impact of abortion decriminalization in states that had previously enacted femicide, equitable divorce asset compensation and unilateral divorce laws. As before, I examine Oaxaca, which decriminalized abortion after implementing these prior legal reforms. I compare municipalities in Oaxaca to those in states that introduced femicide and divorce reforms in the same period but had not yet decriminalized abortion. As shown in Table 3, column 3, there is no statistically significant change in female homicides following abortion decriminalization.

**Table 3**

The difference-in-differences on the number of homicides of women

| | Impact of femicide law | Impact of unilateral divorce in states with femicide law | Impact of decriminalizing abortion in states with femicide, unilateral divorce and equitable divorce asset laws |
|---|---|---|---|
| | (1) | (2) | (3) |
| DID estimate | -0.002 | -0.011 | 0.092 |
| | (0.066) | (0.048) | (3.547) |
| Number of observations x periods | 28582 | 552505 | 9585 |
| Number of switchers x periods | 8117 | 58866 | 8550 |
| Number of months used for post-treatment | 90 | 90 | 14 |
| Controls: unemployment rates of males and females, the number of males killed by firearms, area and fixed effects | Yes | Yes | Yes |
| Treatment | Enacting femicide law, without unilateral divorce and abortion decriminalization | Enacting unilateral divorce in addition to femicide laws | Decriminalizing abortion, after enacting femicide, unilateral divorce laws and equitable divorce asset compensation |
| Comparison | Without femicide laws, unilateral divorce, or abortion decriminalization | With femicide law, but without unilateral divorce or abortion decriminalization yet | With femicide and unilateral divorce laws, each implemented around the same time |

*Notes*: Standard errors, clustered at the state level and bootstrapped, in parentheses. Significance levels *** p<0.01, ** p<0.05, * p<0.1.

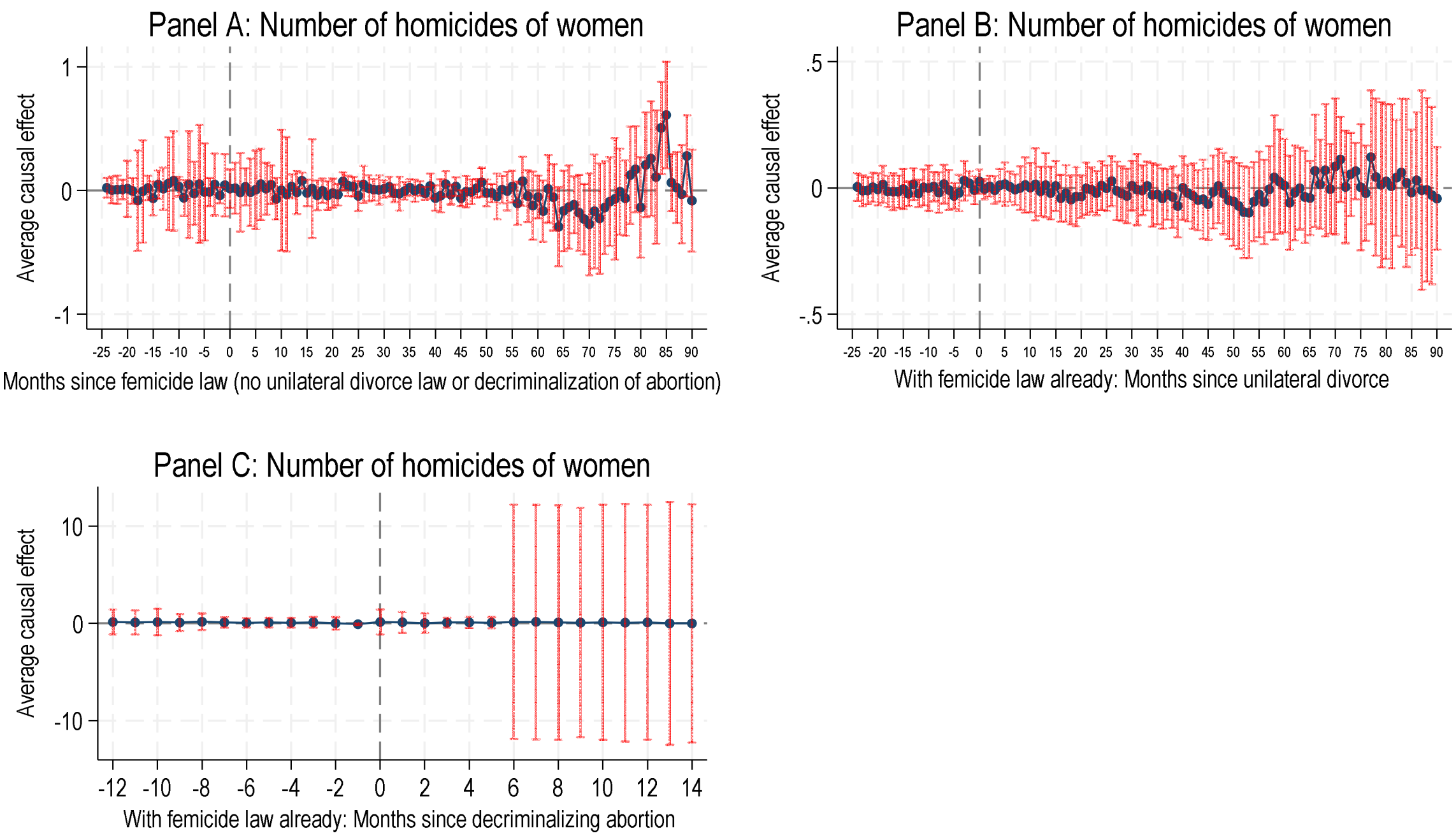

**Fig. 6.** The difference-in-difference on the number of homicides of women

Figure 6 further illustrates the consistent null effects of femicide laws, even when combined with divorce reforms or abortion decriminalization. This figure also reports DID placebo tests, none of which are statistically significant. These results confirm that treatment and control states followed similar pre-treatment homicide trends across all the three legal frameworks analyzed.

### *5.4. Number of disappearances of women*

It is essential to consider that femicide laws may have inadvertently encouraged perpetrators to better conceal their victims' bodies, thus evading arrest and justice. To scrutinize this issue further, Table 4 examines the impact of femicide legislation on reported cases of female disappearances. As mentioned earlier, the government releases these figures based on reports received about people who have disappeared under circumstances likely involving potential crimes.

Following the same approach and control variables, I first examine the effect of femicide laws on the number of disappearances of women. Table 4, column 1, reports the DID estimates, showing no statistically significant effect when comparing states that enacted femicide laws—without concurrent unilateral divorce laws or abortion decriminalization—to similar states that had not yet adopted femicide laws. This is still the case even when using a stricter test, which also excludes the states that at some point implemented equitable divorce asset compensation (see Table A.2, column 3, and Figure A.3, Panel C).

Next, I assess whether the introduction of unilateral divorce laws, after femicide laws were already in place, influenced the number of female disappearances. The estimates in Table 4, column 2, suggest a small but statistically significant effect (DID = 0.09). However, further robustness checks in the next section reveal that this effect is not stable across alternative model specifications, suggesting that the small result found here is not robust.

Finally, I analyze the case of Oaxaca, which decriminalized abortion after enacting femicide laws, unilateral divorce laws and equitable divorce asset compensation. As shown in Table 4, column 3, there is no evidence of a statistically significant impact on the disappearances of women following these legal reforms.

The lack of impact of femicide laws, with or without additional unilateral divorce laws or the decriminalization of abortion, on the number of disappearances of women is also shown in Figure 7. This figure further demonstrates no violation of parallel trends for any of the pre-periods analyzed and no impact of such laws for any of the periods analyzed.

**Table 4**

The difference-in-differences on the number of disappearances of women

| | Impact of femicide law | Impact of unilateral divorce in states with femicide law | Impact of decriminalizing abortion in states with femicide, unilateral divorce and equitable divorce asset laws |
|---|---|---|---|
| | (1) | (2) | (3) |
| DID estimate | 0.015 | 0.094** | 0.342 |
| | (0.012) | (0.037) | (0.437) |
| Number of observations x periods | 28582 | 552505 | 9585 |
| Number of switchers x periods | 8117 | 58866 | 8550 |
| Number of months used for post-treatment | 90 | 90 | 14 |
| Controls: unemployment rates of males and females, the number of males killed by firearms, area and fixed effects | Yes | Yes | Yes |
| Treatment | Enacting femicide law, without unilateral divorce and abortion decriminalization | Enacting unilateral divorce in addition to femicide laws | Decriminalizing abortion, after enacting femicide, unilateral divorce laws and equitable divorce asset compensation |
| Comparison | Without femicide laws, unilateral divorce, or abortion decriminalization | With femicide law, but without unilateral divorce or abortion decriminalization yet | With femicide and unilateral divorce laws, each implemented around the same time |

*Notes*: Standard errors, clustered at the state level and bootstrapped, in parentheses. Significance levels *** p<0.01, ** p<0.05, * p<0.1.

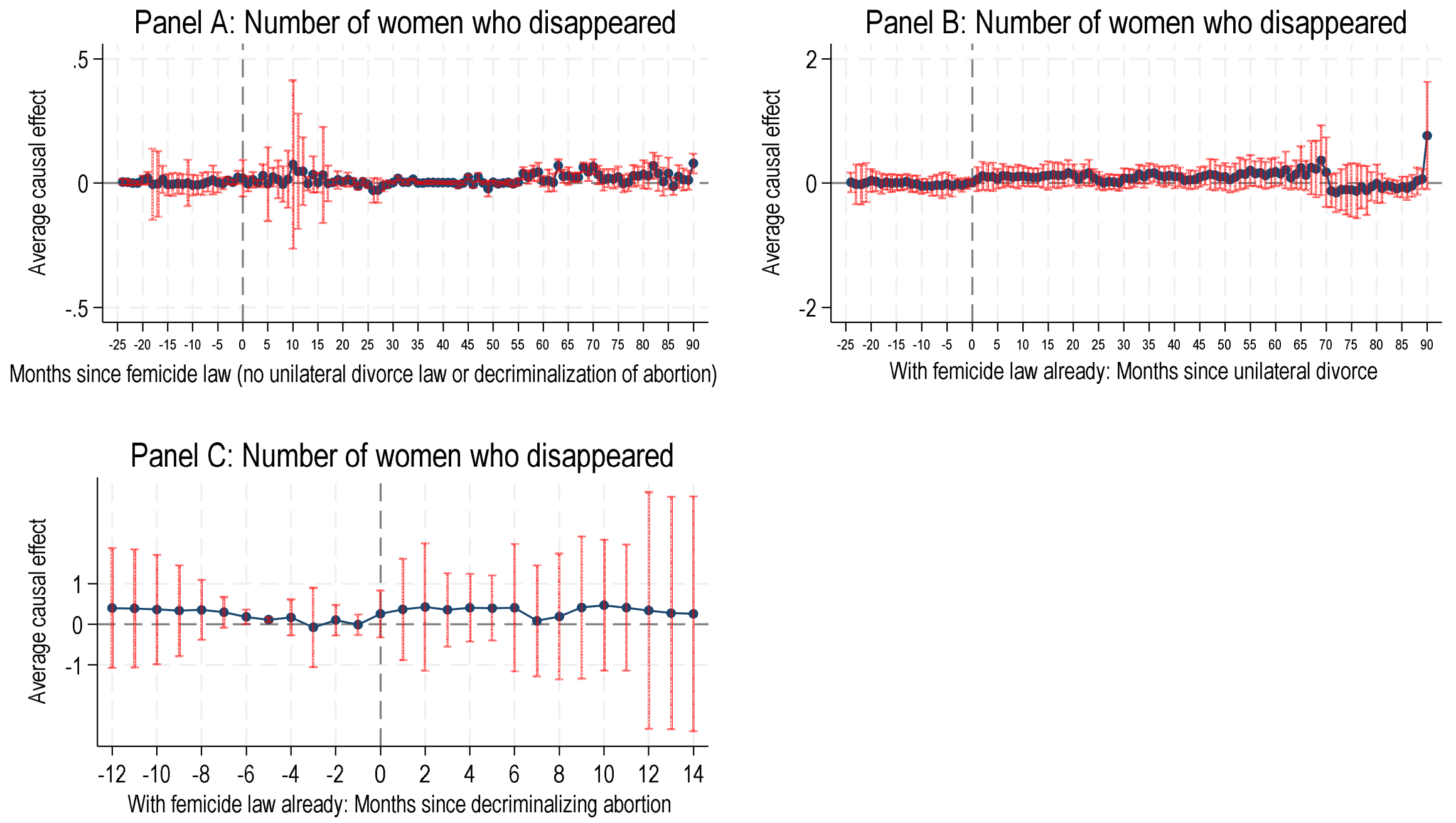

**Fig. 7**. The difference-in-difference on the number of disappearances of women

## *5.5 Suicides*

Another aspect worth examining is the likely impact of femicide laws on female suicides. When penalties for femicide increase, perpetrators may have stronger incentives to conceal gender-based killings by staging them as suicides. A similar pattern may arise following divorce reforms and the decriminalization of abortion. Additionally, changes in the legal landscape could exacerbate psychological distress in abusive relationships, potentially contributing to an increase in female suicides. While experienced forensic specialists may identify staged suicides, a systematic increase in suicides following the law's enactment could indicate wider unintended consequences of these legal reforms.

**Table 5**

The difference-in-differences on the number of suicides of women

| | Impact of femicide law | Impact of unilateral divorce in states with femicide law | Impact of decriminalizing abortion in states with femicide, unilateral divorce and equitable divorce asset laws |
|---|---|---|---|
| | (1) | (2) | (3) |
| DID estimate | -0.036 | 0.006 | 0.037 |
| | (0.037) | (0.011) | (0.963) |
| Number of observations x periods | 28582 | 552505 | 9585 |
| Number of switchers x periods | 8117 | 58866 | 8550 |
| Number of months used for post-treatment | 90 | 90 | 14 |
| Controls: unemployment rates of males and females, the number of males killed by firearms, area and fixed effects | Yes | Yes | Yes |
| Treatment | Enacting femicide law, without unilateral divorce and abortion decriminalization | Enacting unilateral divorce in addition to femicide laws | Decriminalizing abortion, after enacting femicide, unilateral divorce laws and equitable divorce asset compensation |
| Comparison | Without femicide laws, unilateral divorce, or abortion decriminalization | With femicide law, but without unilateral divorce or abortion decriminalization yet | With femicide and unilateral divorce laws, each implemented around the same time |

*Notes*: Standard errors, clustered at the state level and bootstrapped, in parentheses. Significance levels *** $p<0.01$, ** $p<0.05$, * $p<0.1$.

Table 5 presents the results. First, I examine the impact of enacting femicide laws without concurrent unilateral divorce laws or abortion decriminalization. As shown in Table 5, column 1, there is no statistically significant effect on female suicides. This result holds even when excluding states that implemented equitable divorce asset compensation, as seen in Table A.2, column 4 and Figure A.3, Panel C. I then analyze the effects of enacting unilateral divorce laws alongside femicide laws, as well as the additional impact of abortion decriminalization. In neither case do the estimates indicate a statistically significant effect on suicides (Table 5, columns 2 and 3).

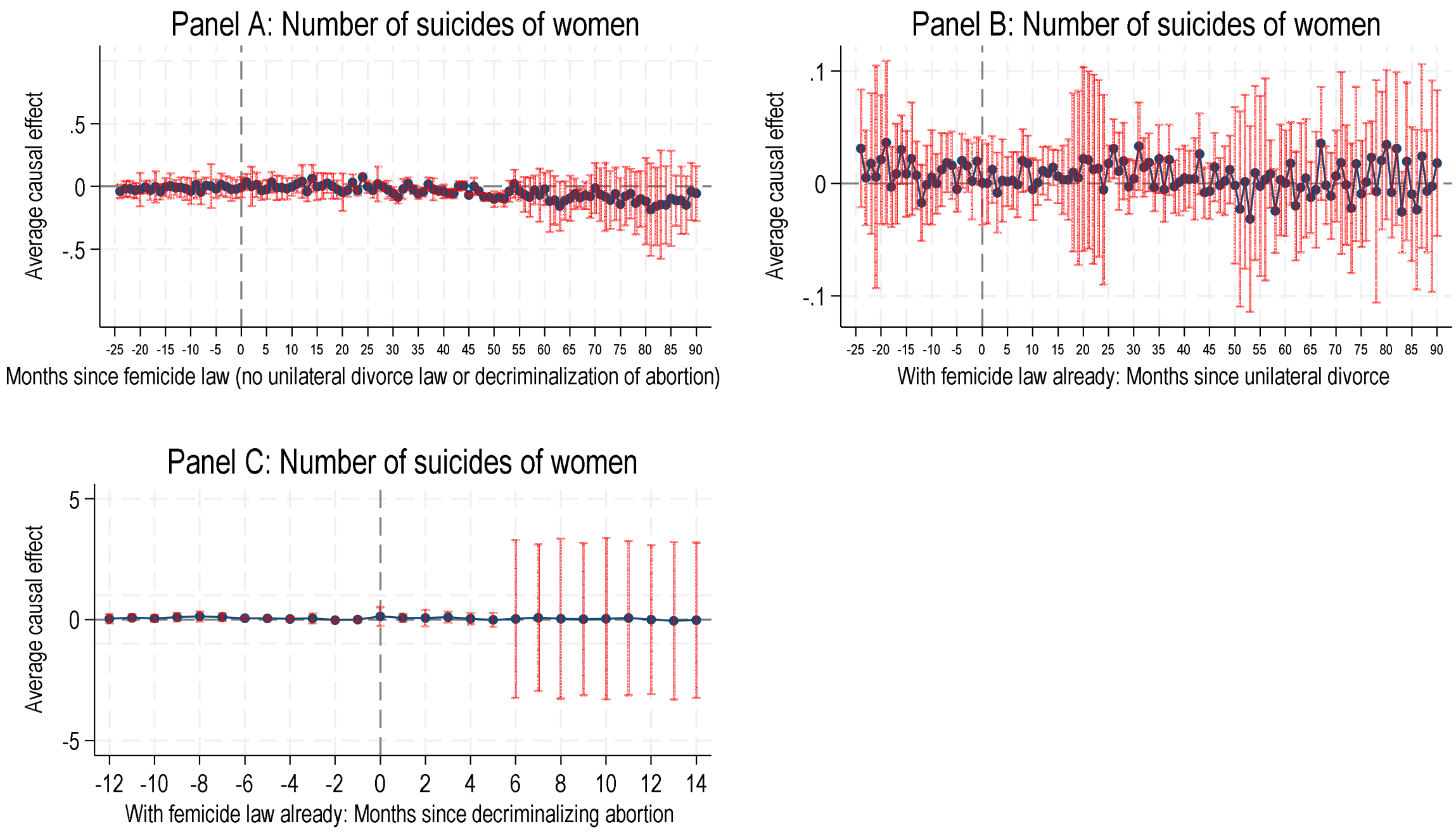

**Fig. 8.** The difference-in-difference on the number of suicides of women

Figure 8 shows there is no statistically significant evidence that the implementation of these laws has led to changes in the number of suicides among women. The placebo tests illustrated in this figure indicate that there are no violations of parallel trends for any of the three laws, reinforcing the suitability of the DID framework employed. While these findings cannot entirely rule out the potential misclassification of homicides as suicides, the results suggest that the enactment of the analyzed laws did not statistically affect the number of suicides among women, as recorded in official death certificates.

## 6. Robustness Checks

I conduct several robustness checks. I start by testing alternative controls. Then, test multiple mechanisms by which femicide laws might influence perpetrators' behavior, including an in-depth analysis of female homicides by 17 subgroups. I also assess differences in prison sentences for femicides across states, and their differential sentences between femicides and homicides. I extend placebo tests to two years prior implementation of femicide laws using alternative specifications, and different measures of lethal violence.

In addition, I analyze divorce reform components, particularly the introduction of new asset compensation for spouses who are dedicated to household and childcare responsibilities. All these tests confirm the null effect of the femicide laws, even when adopted with divorce laws or decriminalization of abortion, on the killing of women. Finally, given the consistency of the DID results, I then use the synthetic control method to further test the robustness of the main findings.

*6.1. Alternative controls*

One potential concern regarding the control variables is that while femicide laws show no impact on female homicide rates, unilateral divorce laws did increase divorce rates. In theory, this could affect female unemployment, though no earlier studies have found such evidence for Mexico (Hoehn-Velasco and Penglase 2021a). To address this, I re-run the analysis with alternative control sets.

I test two alternative control specifications while excluding female unemployment. The first set includes male unemployment (lagged 24 months) and male firearm homicides (lagged 24 months). The lag structure accounts for potential delayed effects of unilateral divorce law on unemployment. The second set of controls removes unemployment and firearm homicides altogether. Instead, it includes a dummy variable indicating whether the state had enacted equitable divorce asset compensation, along with the current annual municipal-level satellite light intensity and its 12-month lag. In both sets of alternative controls, I retain area and time fixed effects. Because some control variables are lagged, the maximum number of post-treatment analysis is reduced for the analyses of unilateral divorce law to 65 periods instead of 90.

The alternative specifications, shown in Table A.3 in the supplementary appendix, confirm that the analyzed reforms had no statistically significant effect on reducing femicides, female homicides, disappearances, or suicides. That is the case whether analyzing states that implemented exclusively femicide laws without any other additional reforms (columns 1-2), those with only equitable divorce asset compensation (columns 3-4), those that combined femicide and unilateral divorce laws (columns 5-6), or those that later decriminalized abortion (columns 7-8).

The dynamic effects of the alternative specifications using the first set of controls are shown in Figure A.4, and those using the second set are shown in Figure A.5. Both figures show no evidence of parallel trends violations and indicate that the analyzed laws had no impact on reducing lethal violence against women in either the short or long run.

*6.2 Mechanisms: homicides by subgroup categories*

The laws analyzed thus far may have had heterogeneous effects on subgroups of women who are particularly more vulnerable to being killed. To address this, I examine whether the impacts of these reforms vary across subgroups defined by Indigenous language use, marital status, schooling attainment, employment status, pregnancy status, and age. This analysis focuses on homicides of women, as defined in official death certificates. Using this broader definition, rather than a particular legal definition of femicide, allows for a more comprehensive examination of killings of women and ensures that all forms of lethal violence against women are captured, regardless of the specific circumstances or motivations.

**Table 6**

The difference-in-differences of the femicide law on female homicide by subgroups

| | DID estimate | Standard error | Number of observations x periods | Number of switchers x periods |
|---|---|---|---|---|
| Speaks indigenous language | 0.001 | (0.001) | 28582 | 8117 |
| Marital status | | | | |
| Married | -0.004 | (0.018) | 28582 | 8117 |
| Divorced | 0.011 | (0.012) | 28582 | 8117 |
| Separated | 0.006 | (0.010) | 28582 | 8117 |
| Educational attainment | | | | |
| Less than primary or not specified | -0.017 | (0.026) | 28582 | 8117 |
| Primary or secondary | 0.022 | (0.025) | 28582 | 8117 |
| High school + | -0.005 | (0.027) | 28582 | 8117 |
| Work status | | | | |
| Worked | 0.005 | (0.011) | 28582 | 8117 |
| Did not work | 0.302 | (0.068) | 239065 | 18557 |
| Pregnancy status | | | | |
| Was pregnant | 0.006 | (0.005) | 28582 | 8117 |
| All pregnancy and post-pregnancy related deaths | 0.012 | (0.011) | 28582 | 8117 |
| Age range | | | | |
| Younger than one year (excluding fetal deaths) | -0.003 | (0.007) | 28582 | 8117 |
| Younger than 14 (excluding fetal deaths) | -0.015 | (0.019) | 28582 | 8117 |
| Aged 15-24 | 0.012 | (0.020) | 28582 | 8117 |
| Aged 25-39 | -0.014 | (0.024) | 28582 | 8117 |
| Aged 40 to 59 | 0.008 | (0.033) | 28582 | 8117 |
| Aged 60 to 99 | 0.006 | (0.006) | 28582 | 8117 |

*Notes*: The treatment group has enacted only the femicide law, without implementing unilateral divorce laws or decriminalizing abortion. The control group, in contrast, has not yet enacted the femicide law. The periods analyzed include 90 months post-treatment. Standard errors, clustered at the state level and bootstrapped, in parentheses.

Significance levels *** $p<0.01$, ** $p<0.05$, * $p<0.1$.

For consistency, the subgroup analyses use the same treatment and control groups, as well as the same pre- and post-treatment periods employed in the main analyses of each law, shown in Section 5. Table 6 presents the results for the femicide law. The treatment group includes states that enacted femicide laws without concurrent unilateral divorce laws or abortion decriminalization, while the control group consists of similar states that had not yet implemented femicide laws. The analysis spans 90 months post-treatment. I use the same controls as in the main analysis.

The DID estimates in Table 6 show no statistically significant impact on female homicides for any of the analyzed groups, including those potentially at higher risk due to socioeconomic or cultural factors. Figure 9 shows no evidence of parallel trends violations before treatment and no indication of a reduced risk of lethal violence against women in the 90 months following treatment.

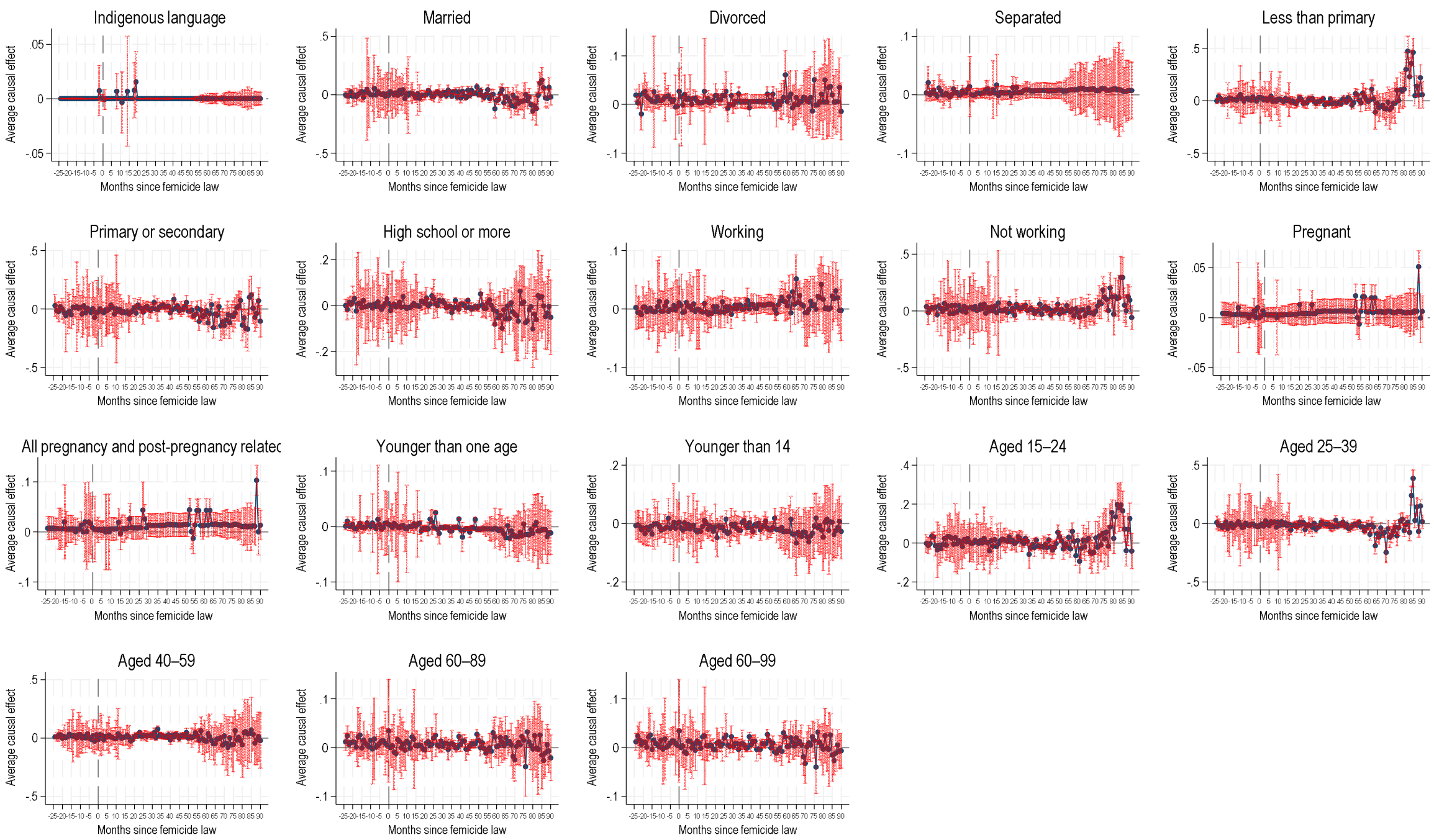

**Fig. 9.** The difference-in-differences of the femicide law on female homicide by subgroups

Then I move on to test whether the introduction of unilateral divorce laws, enacted on top of femicide legislation, had any effect on the number of homicides of women across different subgroups. The treatment group consists of states that implemented both femicide and unilateral divorce laws, while the control group includes those that enacted only femicide laws. The analysis spans 90 months post-treatment. I use the same controls as in Section 5.

The findings in Table A.4 show no statistically significant change in the number of female homicides across any subgroup. The dynamic effects for each subgroup over the 90-month post-treatment period and the 24-month placebo period are presented in Figure A.6. The results show no evidence of parallel trends violations. Notably, there is no evidence that unilateral divorce laws increased homicide risks for married, divorced, or separated women, nor for women with different levels of education or employment status. If concerns existed that expanding legal autonomy for women seeking divorce might provoke a violent backlash, the data offer no support for such claims. At the same time, there is also no evidence that femicide laws have meaningfully reduced women's risk of lethal violence.

Then, I examine the decriminalization of abortion affected the number of homicides of women across different subgroups. The treatment group consists of the state of Oaxaca, which recently decriminalized abortion while also implementing femicide, unilateral divorce laws and equitable divorce asset compensation. The control group includes states that enacted these legal reforms around the same time as Oaxaca but had not yet decriminalized abortion. Again, the analysis spans 14 months post-treatment.

The results, in Table A.5, align with the broader pattern observed across other legal reforms: there is no statistically significant effect on female homicides in any subgroup. This includes pregnant women, post-pregnant women, young women, and those with lower education or unemployment. The dynamic effects for each subgroup over the 14-month post-treatment period and the 12-month placebo period are presented in Figure A.7. The results show no evidence of parallel trends violations.

If there were concerns that abortion decriminalization might trigger a surge in violence or increase the risk for vulnerable women, the results shown provide no basis for such fears. However, there is no evidence that the reforms analyzed have meaningfully reduced women's homicide risks.

### *6.3 Tougher prison sentences for femicides*

A key aspect of evaluating femicide laws is determining whether their impact depends on sentencing severity. An advantage of the staggered DID estimator used is to allow for measuring the effects of varying treatment intensity, enabling me to test whether states with longer sentences experience different outcomes than those with more lenient penalties. As a robustness check, I assess whether femicide laws are more effective when accompanied by harsher sentences. The analysis retains the same control variables used in the main results, Section 5.

To determine the sentencing severity, I compute the mean of the maximum and minimum femicide sentences specified in each state's criminal code. These sentencing lengths change across states and over time, providing an identification strategy that leverages both variations in the timing of femicide law adoption and differences in sentence length.

To ensure a clean comparison, I first focus on states that enacted femicide laws, excluding those that also introduced unilateral divorce or abortion decriminalization. Table A.6 presents the results, showing no statistically significant effect of sentencing severity on the number of femicides, homicides of women, disappearances, or suicides. Figure A.8 confirms the absence of dynamic effects over the 90-month post-treatment period, with no violations of parallel trends 24 months before the femicide law was enacted.

One might question whether the lack of statistical significance arises from focusing solely on states that did not implement complementary legislation, such as unilateral divorce or the decriminalization of abortion. To address this concern, I extend the analysis to evaluate the impact of femicide sentencing severity across all states that enacted femicide laws, irrespective of whether they subsequently adopted additional laws. This evaluation also includes a post-treatment period of 90 months following the enactment of femicide laws. Once again, the findings reveal not statistically significant (Table A.7).

*6.4 Femicides and homicides differential prison sentences*

Potential criminals may not consider the expected average prison sentences for femicides in isolation; they are likely to weigh the additional time they might serve compared to existing sentences for homicides. This is one of the reasons why femicide laws, in theory, implement tougher sentences relative to those for homicides.

The fourth robustness check assesses whether states with larger disparities in average prison sentences between femicides and homicides have any varying impact on femicides, homicides, disappearances or suicides of women. To consider the broadest number of states, the treatment group includes municipalities in states that implemented femicide laws, regardless of whether they have or will implement unilateral divorce laws, equitable divorce asset compensation or decriminalize abortion.

The findings, in Table A.8, show that the femicide laws have been ineffective in reducing the killing of women, even though most states have introduced longer average prison sentence disparities between femicides and homicides. Figure A.9 illustrates that there are no deviations from parallel trends for the four types of crimes analyzed: femicides, homicides of women, disappearances and suicides of women. Additionally, Figure A.9 shows that the dynamic effects are statistically insignificant across all 90 post-treatment periods for each outcome analyzed.

*6.5 Using instead killing rates*

As previously noted, the primary focus is on analyzing the absolute number of various types of killings rather than killing rates. This approach is driven by the fact that population size may fluctuate due to unrelated factors, potentially distorting the net impact on killings. As an additional robustness check, I examine the DID impact of the average prison sentence for femicide on the rates of femicides, female homicides, female disappearances, and suicides of women per 100,000 inhabitants. I use the same control variables as in the primary analysis. The results, in Table A.9, indicate no statistically significant impact on any of these rates.

In a similar vein, I conduct an additional analysis to evaluate the effect of the disparity in average prison sentences between femicides and homicides, again incorporating the same controls. As with the previous results, no statistically significant impact on any killing rate is observed in the 90 post-treatment period (Table A.10).

*6.6 Placebo tests with alternative specifications*

As mentioned earlier, the enactment of femicide laws in Mexico was part of a broader Latin American movement seeking to distinguish femicides from homicides and impose harsher penalties. A critical question, however, is whether some states in Mexico enacted femicide laws in response to their disproportionate increase in female homicides.

To address this concern, I use the pre-treatment period to 24 months before the adoption of femicide laws while maintaining the same controls used in the main results. I classify as treated *all* 32 states that implemented femicide laws (binary treatment), regardless of whether they had also introduced unilateral divorce, equitable divorce asset compensation or decriminalized abortion. I use states that had not yet enacted femicide laws as the control group.

If legislative adoption were in response to local femicide homicide trends, one would expect to see systematic differences in key pre-treatment trends. However, the results show no statistically significant differences in pre-treatment trends for femicides, female homicides, disappearances, or an aggregate measure of all female killings, including suicides of women. Table A.11 and Figure A.10 confirm that there is no statistically significant difference in trends during the 24 months prior to femicide law enactment across all these outcomes.

A limitation of the robustness test just presented is that it considers all states that implemented femicide laws as treated, without accounting for other legal reforms that might have contributed to a reduction in gender-based violence. To refine the analysis, I maintain the same control variables as before. However, I add three additional dummy variables indicating whether a state had already implemented unilateral divorce laws, introduced asset compensation policies for divorcing spouses responsible for household and childcare duties, or decriminalized abortion.

With this alternative specification, shown in Figure 10, I again find no evidence of pre-treatment differences. Additionally, Table A.12, shows the post-treatment dynamics have no impact on the number of killings. This holds across all examined outcomes: the number of femicides, female homicides, female disappearances, and an aggregated measure combining homicides, disappearances, and suicides.

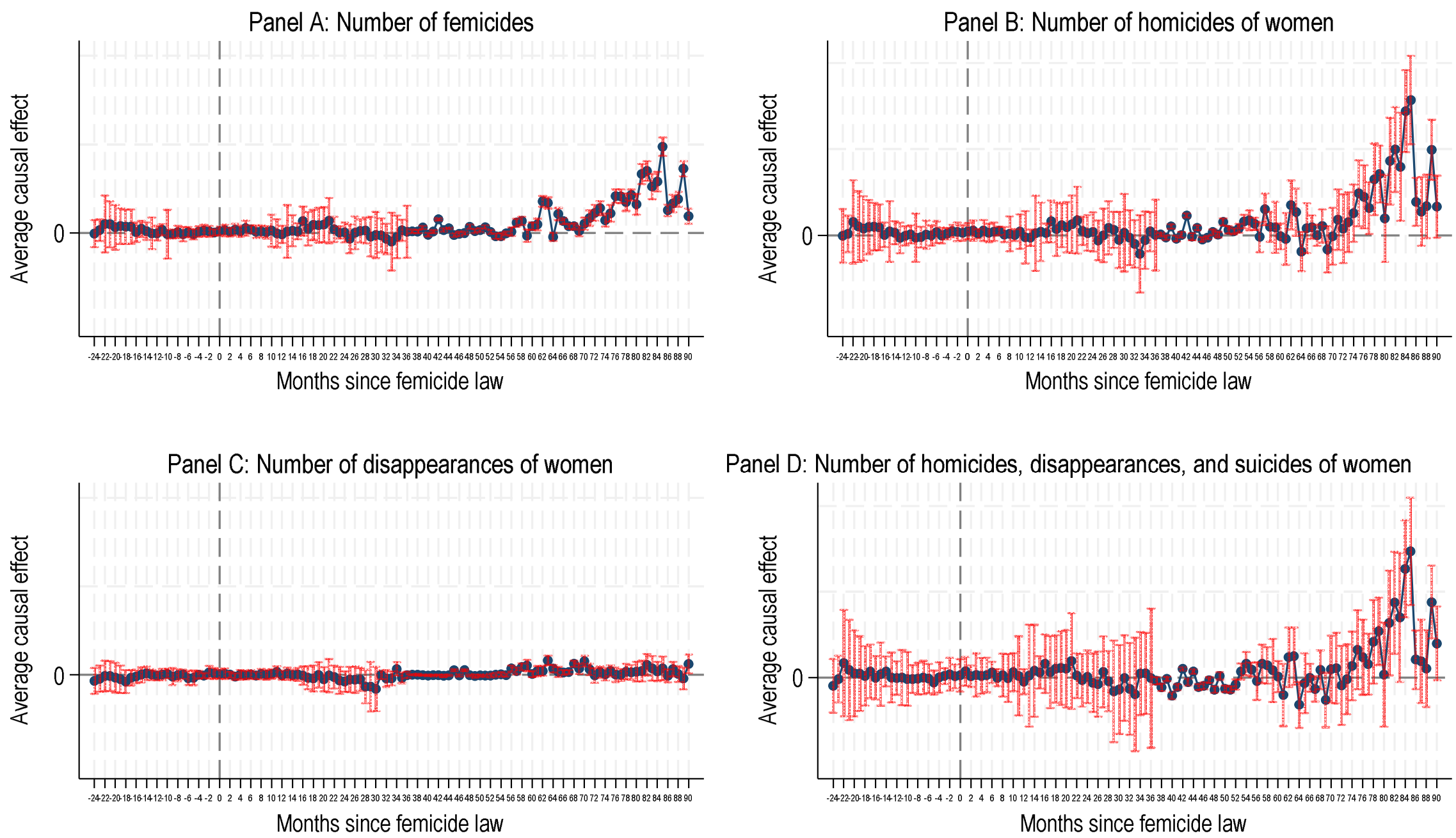

**Fig. 10.** Alternative specification of impact of femicide laws among states that implemented femicide laws, regardless of whether they implemented other laws

### *6.7. Divorce asset compensation and potential mechanisms*

I move on to analyze in far detail the potential impact of the equitable divorce asset compensation. As noted earlier, 23 of Mexico's 32 states gradually introduced financial compensation laws in divorce settlements, requiring the spouse with greater financial assets to transfer up to 50% of marital property to the partner who primarily managed household and childcare responsibilities.[16] These provisions apply exclusively to marriages under separate

[16] Most states introduced equitable divorce asset compensation alongside or soon after unilateral divorce laws, embedding them within broader reforms to marital dissolution. However, there are notable exceptions. Mexico City adopted financial compensation

asset ownership regimes, where prior to the reform, spouses retained only the assets registered in their names.[17] This reform affects a relatively small share of marriages—less than 15%—since most couples in Mexico marry under a communal property regime, which mandates assets acquired during marriage are jointly owned and divided equally upon divorce.

While the asset compensation provisions aim to ensure financial equity, they also create significant liabilities, which may influence divorce behavior and, in extreme cases, trigger unintended consequences such as delayed separations or increased domestic violence. Some individuals might postpone divorce to avoid financial obligations, while others may react violently—whether due to economic loss, frustration, or misinformation.

To examine whether the equitable divorce asset compensation laws led to an increase in lethal violence, I conduct two additional robustness checks using the staggered DID estimator. First, I estimate the impact of equitable divorce asset compensation laws on lethal violence against women over a 90-month period. I compare states that adopted equitable divorce asset compensation reforms with those that have not done so yet. I retain all states in the sample, even if they implemented other potentially relevant reforms, such as femicide laws or unilateral divorce laws. To account for these additional reforms, I include two dummy variables in the controls: one for femicide laws and one for unilateral divorce laws, both of which take the value of one from the month the state adopted such laws onward. I no longer include state-level unemployment or male firearm homicides as controls, as arguably they could have been impacted by the reform. However, I also retain area and time fixed effects as controls.

Figure A.11 confirms that the parallel trends assumption holds in the 24 months preceding implementation. The dynamic estimates show no significant effects on femicides, homicides of women, female disappearances, or suicides, 90 months post implementation.

---

provisions as early as 2000, well before the other reforms on unilateral divorce and femicide legislation.

[17] In Jalisco, the financial compensation provision is capped at a maximum of 40% of the value of assets acquired during the marriage. Despite this limitation, I classify Jalisco as treated under the compensation law framework because the compensation, even at this reduced rate, remains substantial. The compensation is provided alongside other divorce settlement provisions, including custody arrangements, financial support for dependents, spousal maintenance (if applicable), and decisions on who will retain the marital home, if not liquidated, among household items.

Given that women are primary beneficiaries of compensation laws, the inclusion of suicides is particularly relevant, as prolonged exposure to toxic dynamics could exacerbate psychological distress. However, the findings indicate no measurable impact on any of these outcomes (Table A.13).

I also examine the potential effects of financial compensation laws on men. Financial compensation laws, although frequently interpreted through a gendered lens, are fundamentally tied to the roles assumed by spouses during marriage rather than gender itself. Both spouses might perceive the financial obligations imposed by the law as burdensome, potentially creating incentives to avoid these liabilities, sometimes through extreme actions. However, as Figure A.11 illustrates, there is no evidence that the compensation law has contributed to an increase in the homicides of men, disappearances or suicides of men. Table A.13 (columns 5-7) also confirms this result. While national trends for male disappearances and homicides have risen substantially, these increases are not attributable to the divorce asset compensation laws.

The second robustness test examines whether the introduction of unilateral divorce laws altered the impact of equitable divorce asset compensation laws. The control group consists of municipalities in states that had already implemented divorce asset compensation but had not yet introduced unilateral divorce. The treatment group includes municipalities in states where both policies were in place. I include area and time fixed effects as controls, along with a dummy variable indicating whether the state adopted a femicide law. This dummy variable takes a value of one starting from the month the law was adopted, accounting for the potential influence of femicide laws.

Table A.14 confirms that enacting both unilateral and equitable divorce asset compensation laws has no statistically significant impact on any of the examined outcomes for either women or men over the 90-month post-treatment period. Figure A.12 presents the placebo tests for the 24 months before treatment and the 90 months after. The figure clearly shows no statistically significant effect of the divorce reforms on any outcome in any month, whether before or after the reform.

### *6.8 Potential mechanisms in divorce asset compensation*

The equitable divorce asset compensation law aims to reduce economic disparities in divorce settlements but does not appear to affect homicides, disappearances, or suicides. However, its impact may differ for those most vulnerable to violence, particularly individuals in the process of divorce or recently divorced, who may face retaliation from an ex-partner over asset compensation.

While death certificates identify individuals by marital status, they do not indicate the duration of separation or divorce, nor eligibility for asset compensation under a separation of property regime. Despite these limitations, analyzing the impact of divorce asset compensation on homicides among married, divorced, and separated individuals—disaggregated by gender—can offer insights into mechanisms driving gender-based violence. To address potential concerns about confounding variables affecting unemployment or violence. As before, I include area and time fixed effects as controls, along the relevant dummy variables indicating whether femicide laws have been implemented or unilateral divorce laws.

Once again, the results show no statistically significant effects across any of the six outcomes. This conclusion holds regardless of whether the equitable divorce asset compensation law was implemented alongside other laws (Table A.15) or specifically in combination with unilateral divorce laws (Table A.16). The corresponding dynamic effects are shown in Figures A.13 and A.14, which reveal no statistically significant differences in the 24 months before treatment or the 90 months after for any of the six homicide outcomes among married, divorced, and separated individuals, disaggregated by gender.

### *6.9 Synthetic matching difference-in-differences*

Thus far, I have shown that the results of the staggered DID are robust, all indicating null impacts. As a final robustness check, I re-run the main results shown in Section 5 using this time the Synthetic Matching Difference-in-Differences (scDID) method, developed by Arkhangelsky et al. (2021).

This innovative estimator overcomes many of the common pitfalls associated with standard DID approaches. Like other synthetic control methods, scDID constructs an optimally matched control unit, thereby reducing reliance on the strict assumption of parallel trends across aggregates. Crucially, unlike traditional matching methods, scDID does not require that the treated unit lie within the convex hull of the control units—a limitation that often hampers the validity of synthetic controls when the treated unit's characteristics are not well-represented by any combination of control units. Additionally, the scDID framework accommodates multiple treated units, staggered treatment adoption, and the inclusion of covariates.

To maintain comparability with previous analyses, I use the same control variables as the ones used in Section 5 and re-run all estimations for the effects of femicide laws—whether implemented in isolation, alongside unilateral divorce reforms, equitable divorce asset compensation, or in conjunction with the decriminalization of abortion. I analyze femicides, female homicides, disappearances, and suicides.

The findings remain unchanged: there is no statistically significant effect of femicide laws on any of these outcomes (Table 7). While my primary identification strategy relies on staggered DID, scDID serves as an additional check, ensuring that the findings are not driven by poor counterfactual construction. This consistency in results across methodologies reinforces the conclusion that femicide laws have not led to measurable reductions in killings or disappearances of women.

**Table 7**

Synthetic difference-in-differences on the number of femicides, homicides, disappearances, and suicides of women

| | Impact of femicide law without other laws examined | Impact of femicide law without unilateral divorce laws or abortion decriminalization | Impact of unilateral divorce in states with femicide law | Impact of decriminalizing abortion in states with femicide, unilateral divorce and equitable divorce asset laws |
|---|---|---|---|---|
| | (1) | (2) | (3) | (4) |
| Panel A: Number of femicides | | | | |
| scDID estimate | 0.021 | 0.060 | -0.003 | -0.020 |
| | (0.024) | (0.041) | (0.319) | (0.019) |
| Panel B: Number of homicides of women | | | | |
| scDID estimate | 0.029 | 0.134 | 0.000 | 0.020 |
| | (0.030) | (0.100) | (0.418) | (0.025) |
| Panel C: Number of women who disappeared | | | | |
| scDID estimate | -0.006 | -0.010 | 0.039 | -0.021 |
| | (0.010) | (0.011) | (0.308) | (0.141) |
| Panel D: Number of suicides of women | | | | |
| scDID estimate | 0.033 | -0.008 | 0.001 | -0.002 |
| | (0.040) | (0.013) | (0.015) | (0.018) |
| Number of observations x periods | 8,383 | 36,740 | 149,770 | 19,170 |
| Controls: unemployment rates of males and females, the number of males killed by firearms, area and fixed effects | Yes | Yes | Yes | Yes |
| Treatment | Enacted femicide law but has no unilateral divorce law, divorce asset compensation, or abortion decriminalization | Enacting femicide law, without unilateral divorce and abortion decriminalization | Enacting unilateral divorce in addition to femicide laws | Decriminalizing abortion, after enacting femicide, unilateral divorce laws and equitable divorce asset compensation |
| Comparison | Without femicide laws, unilateral divorce, equitable asset compensation, or abortion decriminalization | Without femicide laws, unilateral divorce, or abortion decriminalization | With femicide law, but without unilateral divorce or abortion decriminalization yet | With femicide and unilateral divorce laws, each implemented around the same time |

*Notes*: Standard errors, clustered at the state level and bootstrapped, in parentheses. Significance levels *** p<0.01, ** p<0.05, * p<0.1.

## 7. Conclusion

This paper demonstrated that femicide laws in Mexico have failed to reduce femicides, homicides of women, disappearances, and suicides of women. This holds true regardless of whether states imposed harsher prison sentences for femicides compared to homicides and whether they introduced additional reforms such as unilateral divorce, equitable divorce asset compensation, or abortion decriminalization. In essence, the femicide laws have not yet provided the intended protection for women in either the short term or even after 90 months post-enactment.

Before the adoption of femicide laws in Latin America, a range of international legislative measures had already been enacted to combat gender-based and intimate partner violence. For instance, research in the US shows that arresting individuals involved in intimate partner violence yields mixed results, depending on whether arrests are mandated or left to police discretion (Chin and Cunningham 2019; Iyengar 2009). Mexico has not implemented such mandates, and given its weak judicial capacity and high rates of impunity, the effectiveness of such measures is doubtful. Criminalizing domestic violence has been another widely adopted approach (Dugan, Rosenfeld, and Nagin 2003; Beleche 2019). Long before the period of analysis, Mexico had enacted such laws nationwide in the mid-1990s. Evidence suggests these laws reduced sexual and physical violence among married women (Beleche 2019). However, the subsequent rise in homicides analyzed in this paper occurred during a period marked by generalized violence, widespread availability of weapons, and rampant lawlessness, rendering lives alarmingly disposable.

Overall, the findings hold important implications for countries that have recently adopted or are considering adopting femicide legislation. The central lesson is that recognizing femicides as hate crimes and imposing harsh sentences, while important, does not effectively deter perpetrators in settings characterized by widespread impunity.

## Online appendix

**Table A.1**

Date of enactment of femicide, unilateral divorce laws, equitable divorce asset compensation and decriminalization of abortion, 2000-2020

| State | Aggravated homicide due to gender reasons | Femicide | Unilateral divorce | 50% division of assets separate property regime | Abortion |
|---|---|---|---|---|---|
| Aguascalientes | 20/05/2013 | 21/08/2017 | 22/06/2015 | 22/06/2015 | |
| Baja California | | 05/06/2012 | | 19/11/2007 | |
| Baja California Sur | 01/02/2014 | 10/04/2019 | 31/12/2016 | | |
| Campeche | | 20/07/2012 | | | |
| Coahuila | | 20/11/2012 | 05/04/2013 | 05/04/2013 | |
| Colima | | 27/08/2011 | 19/03/2016 | 26/05/2007 | |
| Chiapas | | 09/02/2012 | 23/01/2019 | 23/01/2019 | |
| Chihuahua | 27/08/2003 | 28/10/2017 | | | |
| Mexico City | | 26/07/2011 | 03/10/2008 | 25/05/2000 | 24/04/2007 |
| Durango | 11/12/2011 | 25/12/2018 | 19/07/2018 | 01/07/2018 | |
| Guanajuato | | 03/06/2011 | | 27/03/2009 | |
| Guerrero | | 21/12/2010 | 09/03/2012 | 21/12/2010 | |
| Hidalgo | | 01/04/2013 | 31/03/2011 | | |
| Jalisco | | 22/09/2012 | 17/11/2018 | 25/11/2014 | |
| Estado de México | | 18/03/2011 | 03/05/2012 | 06/03/2010 | |
| Michoacán | 21/01/2014 | 21/03/2017 | 30/09/2015 | | |
| Morelos | | 01/09/2011 | 10/03/2016 | 10/03/2016 | |
| Nayarit | 29/09/2012 | 30/09/2016 | 27/05/2015 | 05/05/2007 | |
| Nuevo León | 26/06/2013 | 05/05/2017 | 14/12/2016 | 28/05/2018 | |
| Oaxaca | | 04/10/2012 | 12/05/2017 | 12/05/2017 | 25/09/2019 |
| Puebla | | 31/12/2012 | 29/03/2016 | 29/03/2016 | |
| Querétaro | | 12/06/2013 | 30/11/2016 | 30/11/2016 | |
| Quintana Roo | | 30/05/2012 | 04/07/2017 | 04/07/2017 | |
| San Luis Potosí | | 23/07/2011 | 30/05/2017 | 30/05/2017 | |
| Sinaloa | | 25/04/2012 | 06/02/2013 | 06/02/2013 | |
| Sonora | | 28/11/2013 | | | |
| Tabasco | | 24/03/2012 | | | |
| Tamaulipas | | 22/06/2011 | 14/07/2015 | 14/07/2015 | |
| Tlaxcala | | 09/03/2012 | 10/02/2016 | | |
| Veracruz | | 29/08/2011 | 10/06/2020 | 10/06/2020 | |
| Yucatán | | 11/09/2012 | 30/04/2012 | 30/04/2012 | |
| Zacatecas | | 04/08/2012 | 13/09/2017 | | |

*Notes:* Own search using the criminal, civil, and family law codes.

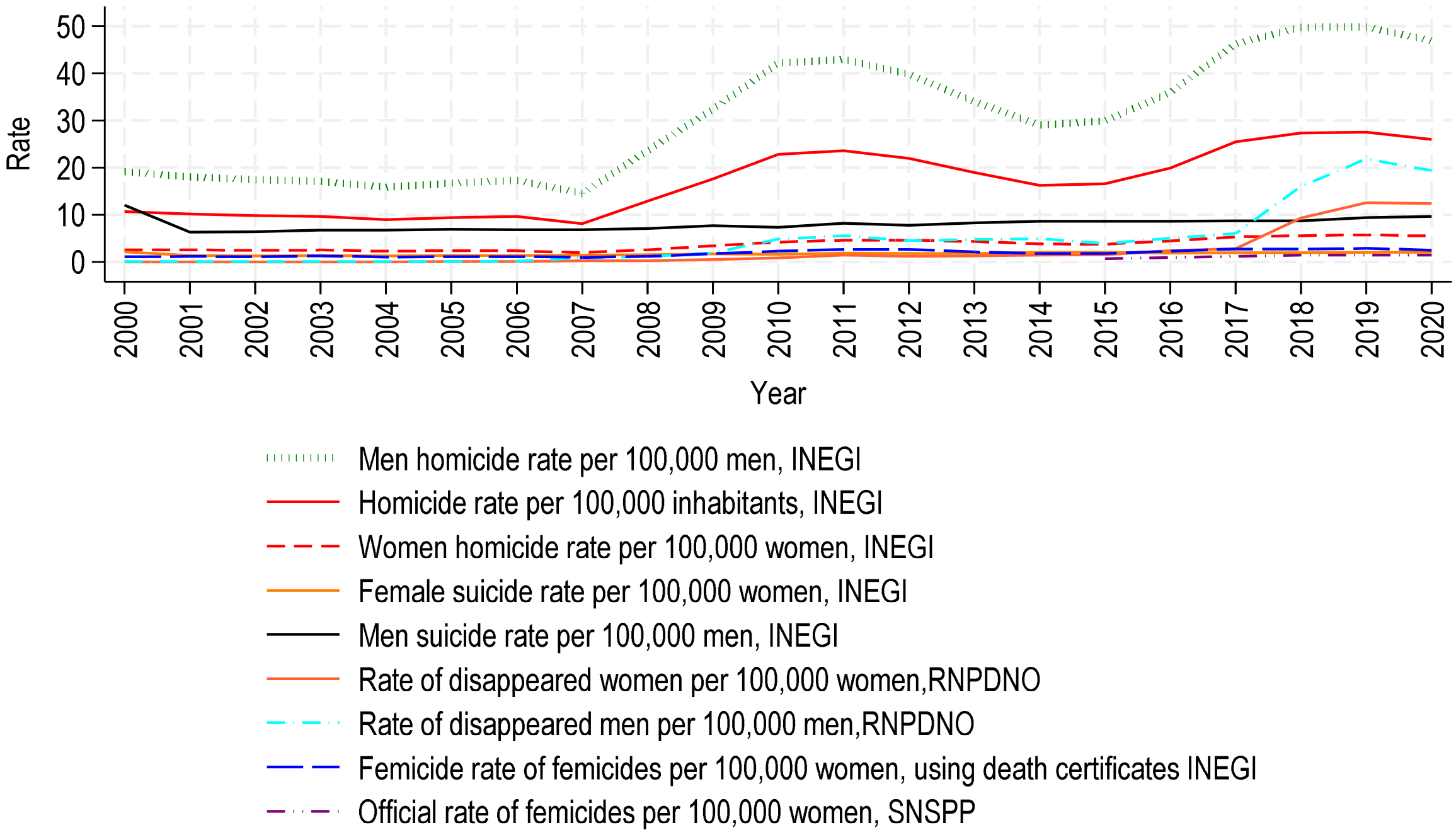

**Fig. A.1.** Rate of homicides, femicides, disappearances and suicides

*Notes:* Homicide, femicide and suicide rates are calculated using death certificates available from the National Institute of Statistics and Geography (INEGI). The rate of disappeared people is sourced from *Registro Nacional de Personas Desaparecidas y No Localizadas* (RNPDNO). The population sizes for women and men used to calculate their respective rates are obtained from the National Population Council (CONAPO).

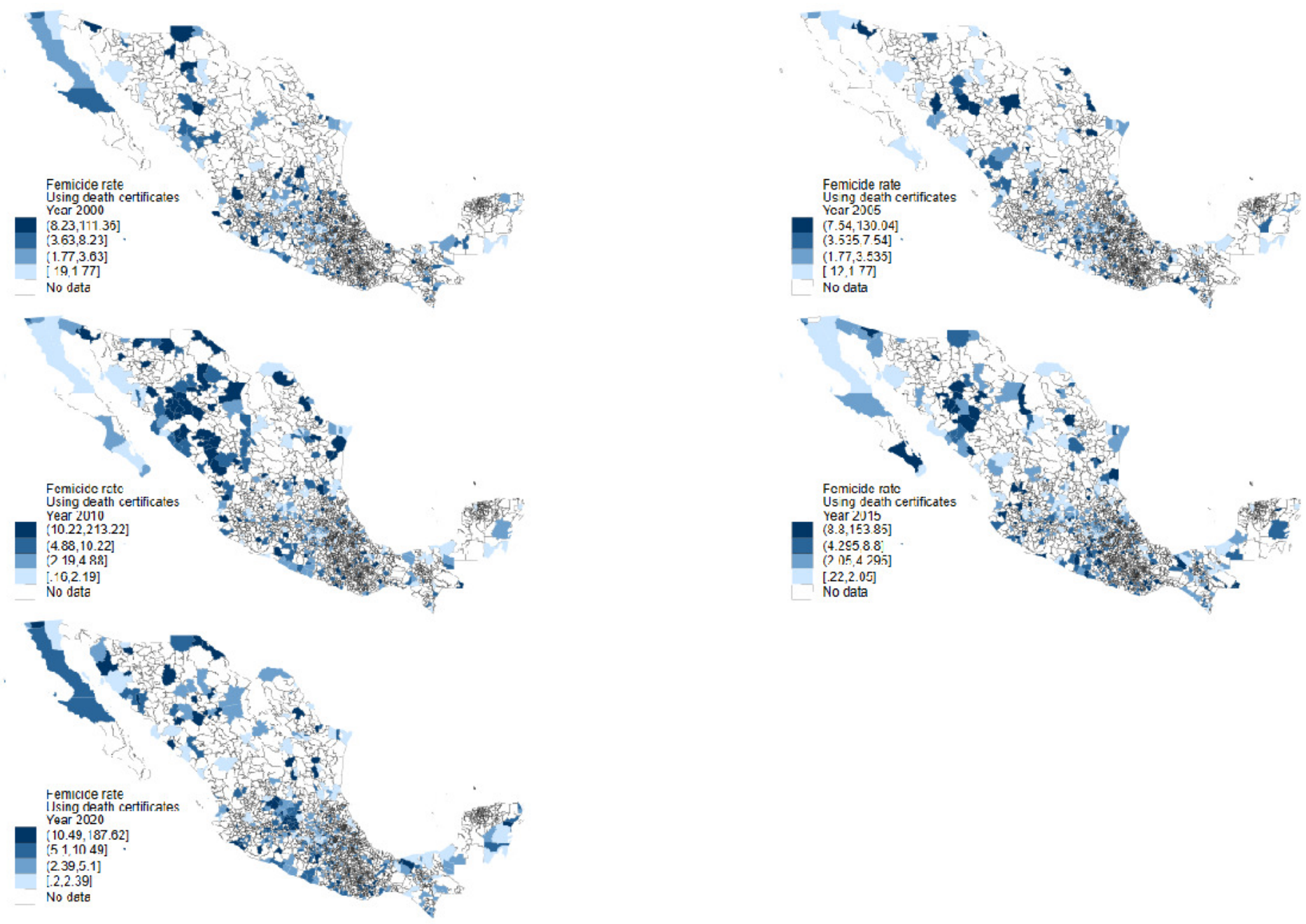

**Fig. A.2.** Femicide rates, per 100,000 women

*Notes:* Femicide rates are calculated using death certificates available from the National Institute of Statistics and Geography (INEGI). The population sizes for women to calculate the femicide rates are obtained from the National Population Council (CONAPO).

**Table A.2**

The difference-in-differences of the femicide law (enacted without unilateral divorce law, equitable divorce asset compensation, or abortion decriminalization)

| | Number of femicides | Number of homicides of | Number of women who disappeared | Number of suicides of women |
|---|---|---|---|---|
| | (1) | (2) | (3) | (4) |
| DID estimate | 0.035** | 0.006 | 0.027 | 0.009 |
| | (0.018) | (0.019) | (0.032) | (0.044) |
| Number of observations x periods | 2816 | 2816 | 2816 | 2816 |
| Number of switchers x periods | 464 | 464 | 464 | 464 |
| Number of months used for post-treatment | 15 | 15 | 15 | 15 |
| Controls: unemployment rates of males and females, the number of males killed by firearms, area and fixed effects | Yes | Yes | Yes | Yes |
| Treatment | Enacting femicide law, without unilateral divorce, equitable divorce asset | | | |
| Comparison | Without femicide laws, unilateral divorce, equitable divorce asset compensation, or abortion decriminalization | | | |

*Notes*: Standard errors, clustered at the state level and bootstrapped, in parentheses. Significance levels *** p<0.01, ** p<0.05, * p<0.1.

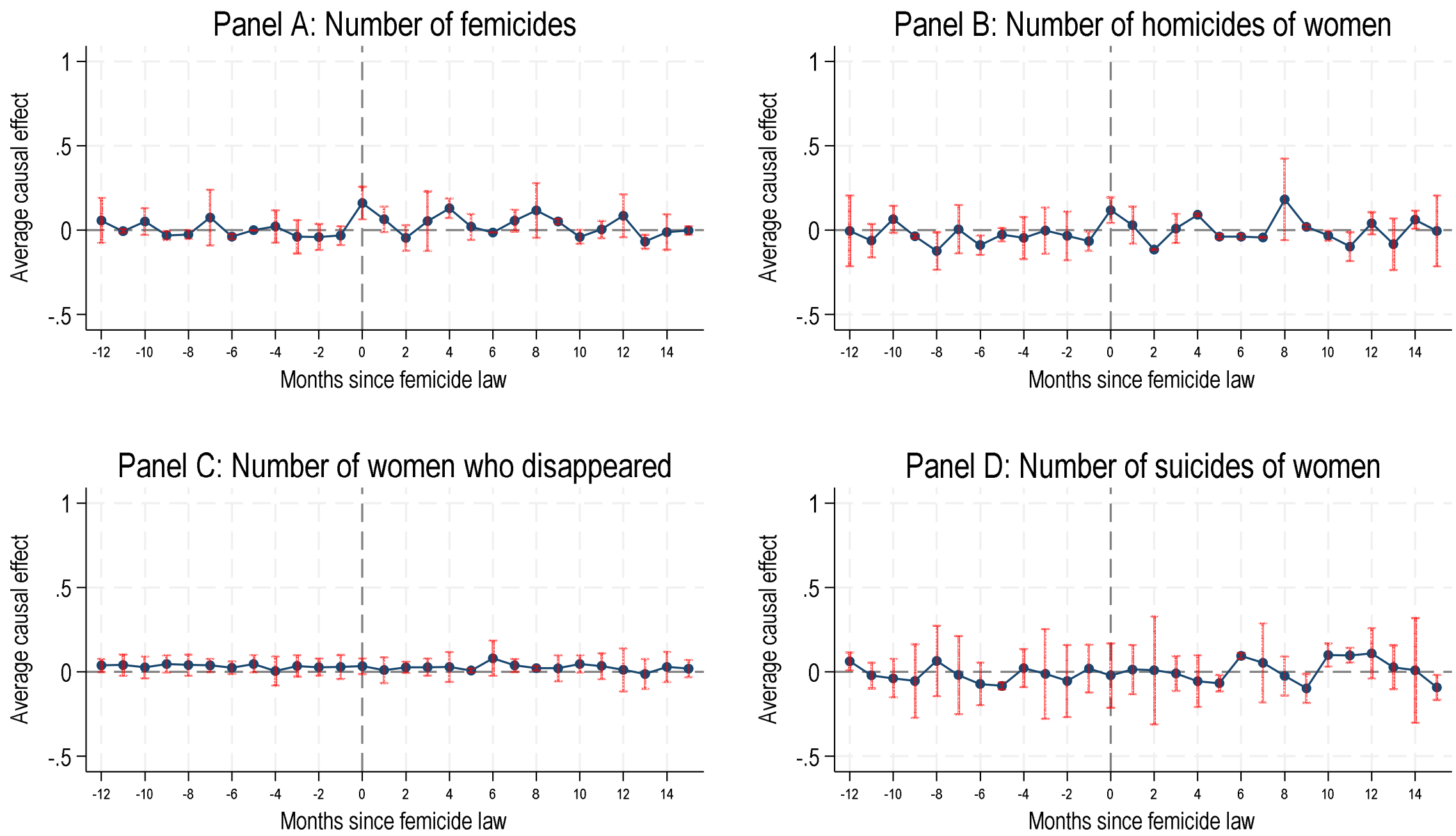

**Fig. A.3.** The difference-in-differences of the femicide law (enacted without unilateral divorce law, equitable divorce asset compensation, or abortion decriminalization)

## Table A.3

The difference-in-differences on the number of femicides, homicides, disappearances, and suicides of women, using alternative controls

| | Impact of femicide law without other laws examined | Impact of femicide law without other laws examined | Impact of femicide law without unilateral or abortion decriminalization | Impact of femicide law without unilateral or abortion decriminalization | Impact of unilateral divorce in states with femicide law | Impact of unilateral divorce in states with femicide law | Impact of decriminalizing abortion in states with femicide, unilateral divorce and equitable divorce asset laws | Impact of decriminalizing abortion in states with femicide, unilateral divorce and equitable divorce asset laws |
|---|---|---|---|---|---|---|---|---|
| | (1) | (2) | (3) | (4) | (5) | (6) | (7) | (8) |
| Panel A: Number of femicides | | | | | | | | |
| DID estimate | 0.031** | 0.032* | 0.077 | 0.033 | -0.051 | -0.010 | -0.059 | -0.024 |
| | (0.016) | (0.018) | (0.040) | (0.048) | (0.036) | (0.052) | (0.088) | (0.274) |
| Panel B: Homicides of women | | | | | | | | |
| DID estimate | 0.002 | 0.003 | 0.107 | 0.011 | -0.080 | -0.019 | -0.052 | -0.228 |
| | (0.017) | (0.016) | (0.069) | (0.105) | (0.053) | (0.111) | (0.027) | (0.350) |
| Panel C: Women who disappeared | | | | | | | | |
| DID estimate | 0.028 | 0.030 | 0.011 | 0.012 | 0.118 | 0.159 | 0.145 | 0.108 |
| | (0.028) | (0.028) | (0.020) | (0.023) | (0.077) | (0.113) | (0.250) | (0.353) |
| Panel D: Suicides of women | | | | | | | | |
| DID estimate | 0.014 | 0.015 | -0.013 | -0.028 | 0.007 | 0.013 | -0.035 | -0.082 |
| | (0.038) | (0 .040) | (0.013) | (0.024) | (0.012) | (0.025) | (0.050) | (0.212) |
| Number of observations x periods | 3172 | 3152 | 28582 | 28332 | 28582 | 312756 | 9585 | 9585 |
| Number of switchers x periods | 532 | 516 | 8117 | 8083 | 8117 | 48423 | 8550 | 8550 |
| Number of months used for post-treatment | 19 | 19 | 90 | 90 | 65 | 65 | 14 | 14 |
| Controls: | | | | | | | | |
| Unemployment rate of males (lagged by 24 months), number of males killed by firearms (lagged by 24 months), male suicides | Yes | | Yes | | Yes | | Yes | |
| Whether the state had new divorce asset compensation, total annual satellite light (lagged by 12 months), total annual satellite light | | Yes | | Yes | | Yes | | Yes |
| Area and fixed effects | Yes | Yes | Yes | Yes | Yes | Yes | Yes | Yes |
| | Enacted femicide law but has no unilateral divorce law, divorce asset compensation, or abortion decriminalization | | Enacting femicide law, without unilateral divorce and abortion decriminalization | | Enacting unilateral divorce in addition to femicide laws | | Decriminalizing abortion, after enacting femicide, unilateral divorce laws and equitable divorce asset compensation | |
| Comparison | Without femicide laws, unilateral divorce, equitable asset compensation, or abortion decriminalization | | Without femicide laws, unilateral divorce, or abortion decriminalization | | With femicide law, but without unilateral divorce or abortion decriminalization yet | | With femicide and unilateral divorce laws, each implemented around the same time | |

*Notes*: Standard errors, clustered at the state level and bootstrapped, in parentheses.

Significance levels *** p<0.01, ** p<0.05, * p<0.1.

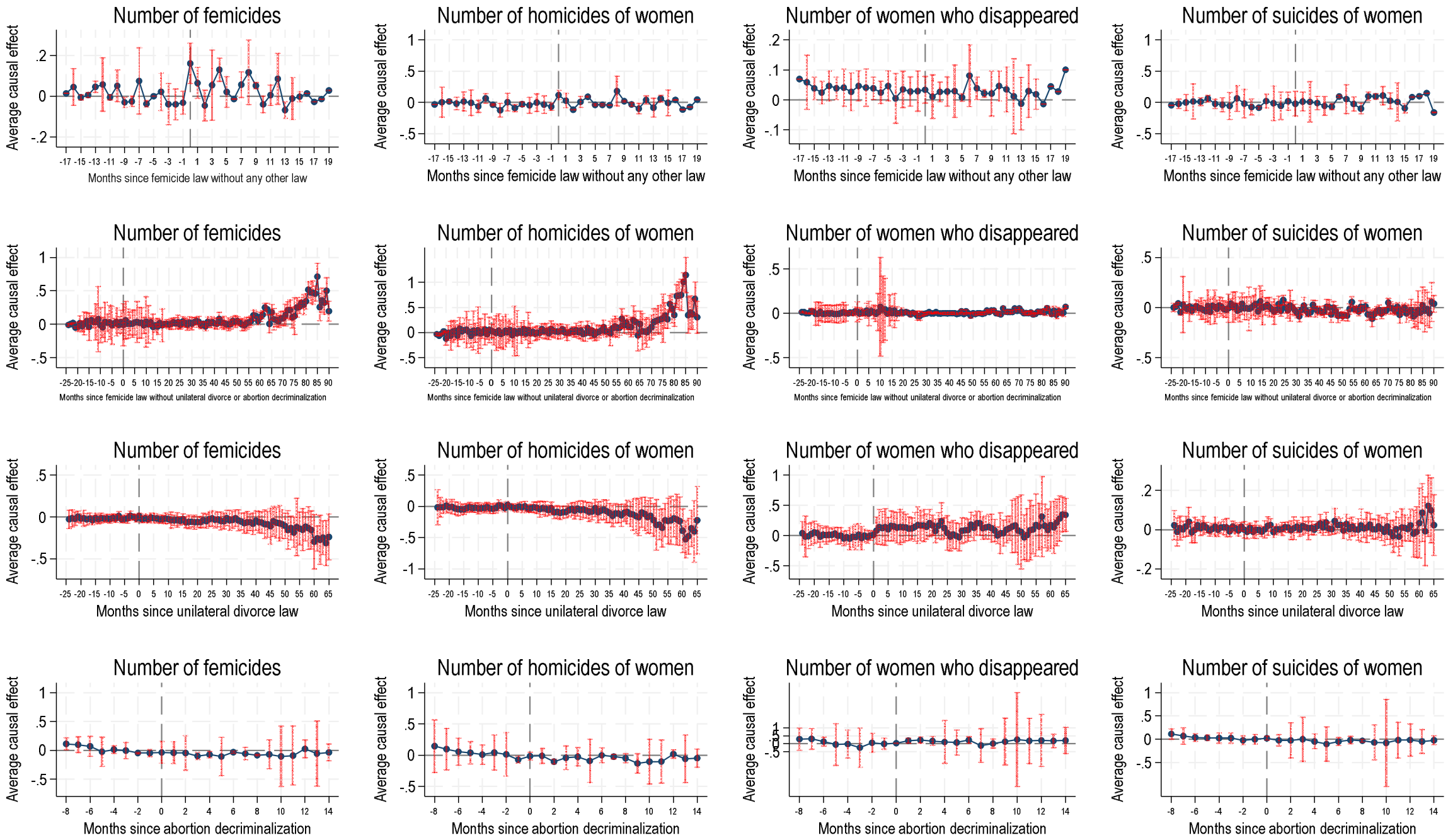

**Fig. A.4.** The difference-in-differences on the number of femicides, homicides, disappearances, and suicides of women, using first set of alternative controls

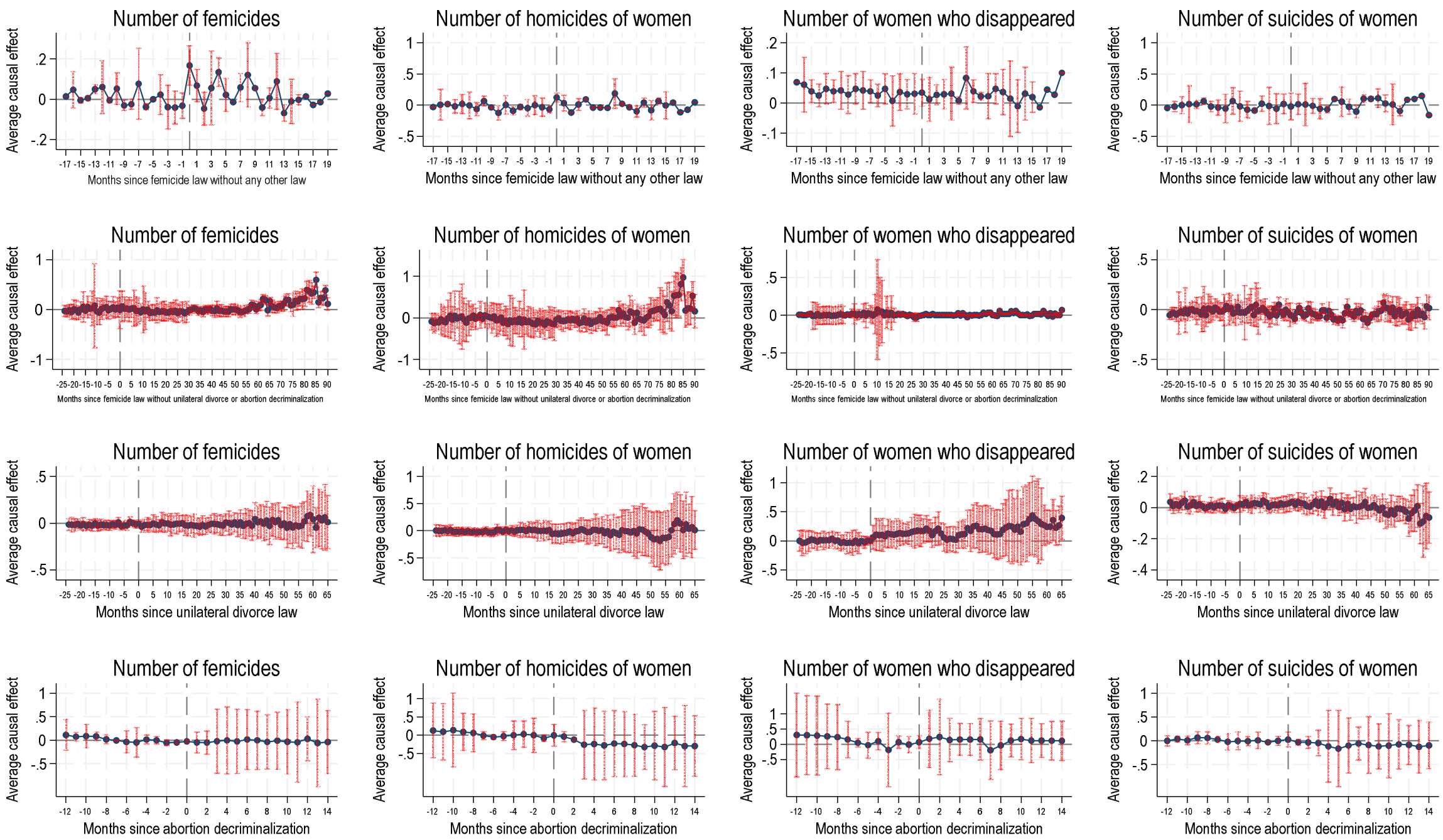

**Fig. A.5.** The difference-in-differences on the number of femicides, homicides, disappearances, and suicides of women, using second set of alternative controls

## Potential mechanisms: analysis by subgroups

**Table A.4**

The difference-in-differences of the unilateral divorce law on homicides of women by subgroups

| | DID estimate | Standard error | Number of observations x periods | Number of switchers x periods |
|---|---|---|---|---|
| Speaks indigenous language | 0.004 | (0.003) | 552505 | 58866 |
| | Marital status | | | |
| Married | -0.015 | (0.028) | 552505 | 58866 |
| Divorced | 0.002 | (0.005) | 552505 | 58866 |
| Separated | -0.003 | (0.003) | 552505 | 58866 |
| | Educational attainment | | | |
| Less than primary or not specified | 0.006 | (0.008) | 552505 | 58866 |
| Primary or secondary | -0.012 | (0.019) | 552505 | 58866 |
| High school + | -0.009 | (0.016) | 552505 | 58866 |
| | Work status | | | |
| Worked | -0.006 | (0.005) | 552505 | 58866 |
| Did not work | -0.015 | (0.033) | 552505 | 58866 |
| | Pregnancy status | | | |
| Was pregnant | 0.001 | (0.002) | 552505 | 58866 |
| All pregnancy and post-pregnancy related deaths | 0.001 | (0.006) | 552505 | 58866 |
| | Age range | | | |
| Younger than one year (excluding fetal deaths) | 0.003 | (0.002) | 552505 | 58866 |
| Younger than 14 (excluding fetal deaths) | 0.003 | (0.004) | 552505 | 58866 |
| Aged 15-24 | -0.013 | (0.012) | 552505 | 58866 |
| Aged 25-39 | -0.022 | (0.026) | 552505 | 58866 |
| Aged 40 to 59 | 0.007 | (0.014) | 552505 | 58866 |
| Aged 60 to 99 | 0.010 | (0.005) | 552505 | 58866 |

*Notes*: The treatment group has enacted both the femicide law and unilateral divorce laws. In contrast, the control group has enacted the femicide law but has not yet implemented unilateral divorce laws. The analysis covers 90 months post-treatment. Standard errors, clustered at the state level and bootstrapped, in parentheses.

Significance levels *** p<0.01, ** p<0.05, * p<0.1.

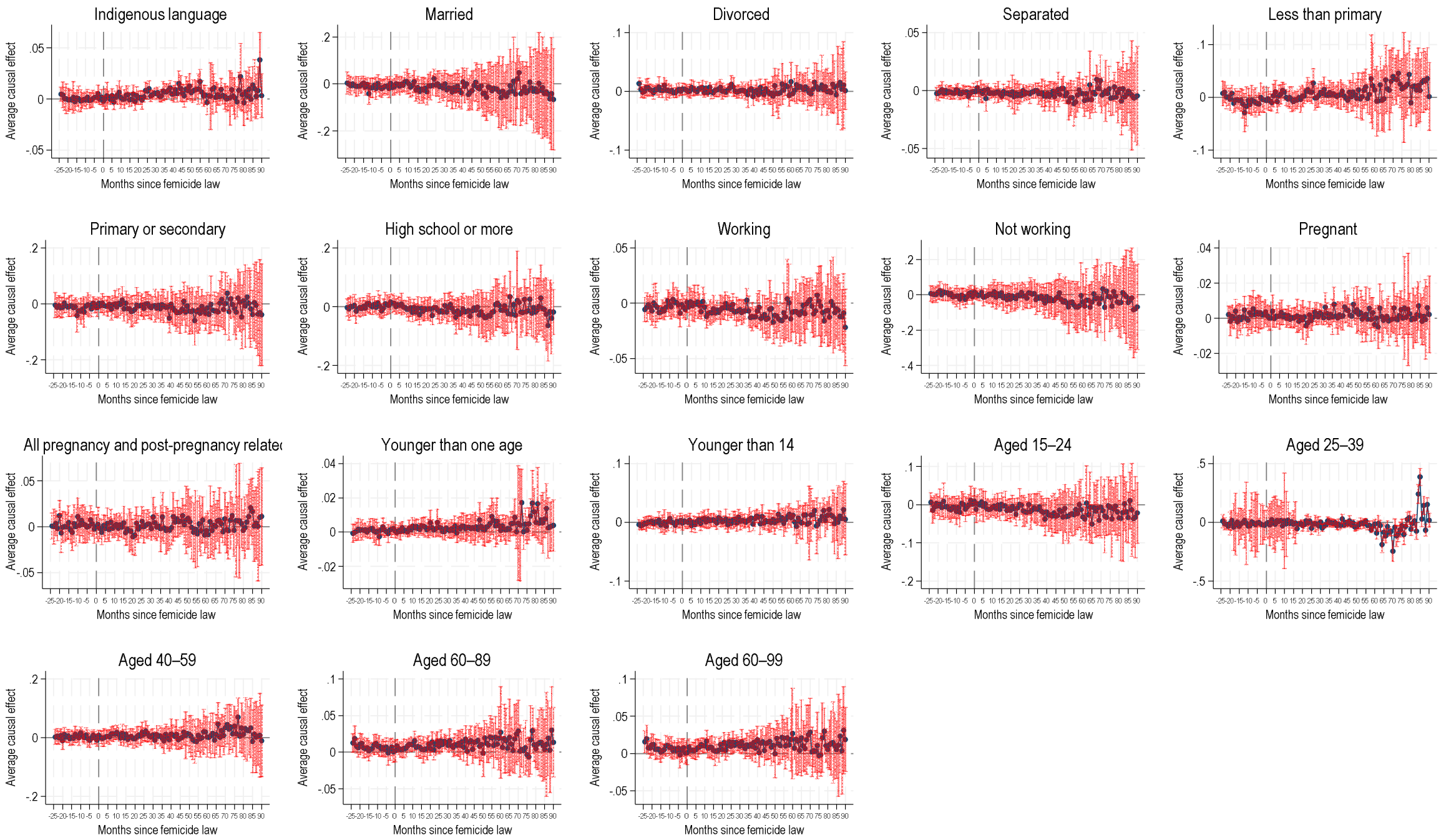

**Fig. A.6.** The difference-in-differences of the unilateral divorce law on homicides of women by subgroups

**Table A.5**

The difference-in-differences of the decriminalization of abortion on homicides of women by subgroups

| | DID estimate | Standard error | Number of observations x periods | Number of switchers x periods |
|---|---|---|---|---|
| Speaks indigenous language | -0.001 | (0.347) | 9585 | 8550 |
| | Marital status | | | |
| Married | 0.063 | (1.848) | 9585 | 8550 |
| Divorced | 0.009 | (0.067) | 9585 | 8550 |
| Separated | -0.005 | (0.006) | 9585 | 8550 |
| | Educational attainment | | | |
| Less than primary or not specified | 0.003 | (0.041) | 9585 | 8550 |
| Primary or secondary | 0.036 | (0.355) | 9585 | 8550 |
| High school + | 0.031 | (2.342) | 9585 | 8550 |
| | Work status | | | |
| Worked | 0.001 | (0.438) | 9585 | 8550 |
| Did not work | 0.072 | (0.224) | 9585 | 8550 |
| | Pregnancy status | | | |
| Was pregnant | -0.016 | (0.486) | 9585 | 8550 |
| All pregnancy and post-pregnancy related deaths | -0.035 | (1.045) | 9585 | 8550 |
| | Age range | | | |
| Younger than one year (excluding fetal deaths) | 0.026 | (0.039) | 9585 | 8550 |
| Younger than 14 (excluding fetal deaths) | 0.029 | (0.246) | 9585 | 8550 |
| Aged 1 to 14 | 0.003 | (0.208) | 9585 | 8550 |
| Aged 15-24 | -0.001 | (0.064) | 9585 | 8550 |
| Aged 25-39 | 0.048 | (2.735) | 9585 | 8550 |
| Aged 40 to 59 | 0.045 | (1.066) | 9585 | 8550 |
| Aged 60 to 99 | -0.029 | (0.072) | 9585 | 8550 |

*Notes*: The treatment group has decriminalized abortion in addition to enacting femicide, unilateral divorce laws and equitable divorce asset compensation. In contrast, the control group has enacted femicide and unilateral divorce laws but has not yet decriminalized abortion. The analysis spans 14 months post-treatment. Standard errors, clustered at the state level and bootstrapped, in parentheses.

Significance levels *** p<0.01, ** p<0.05, * p<0.1.

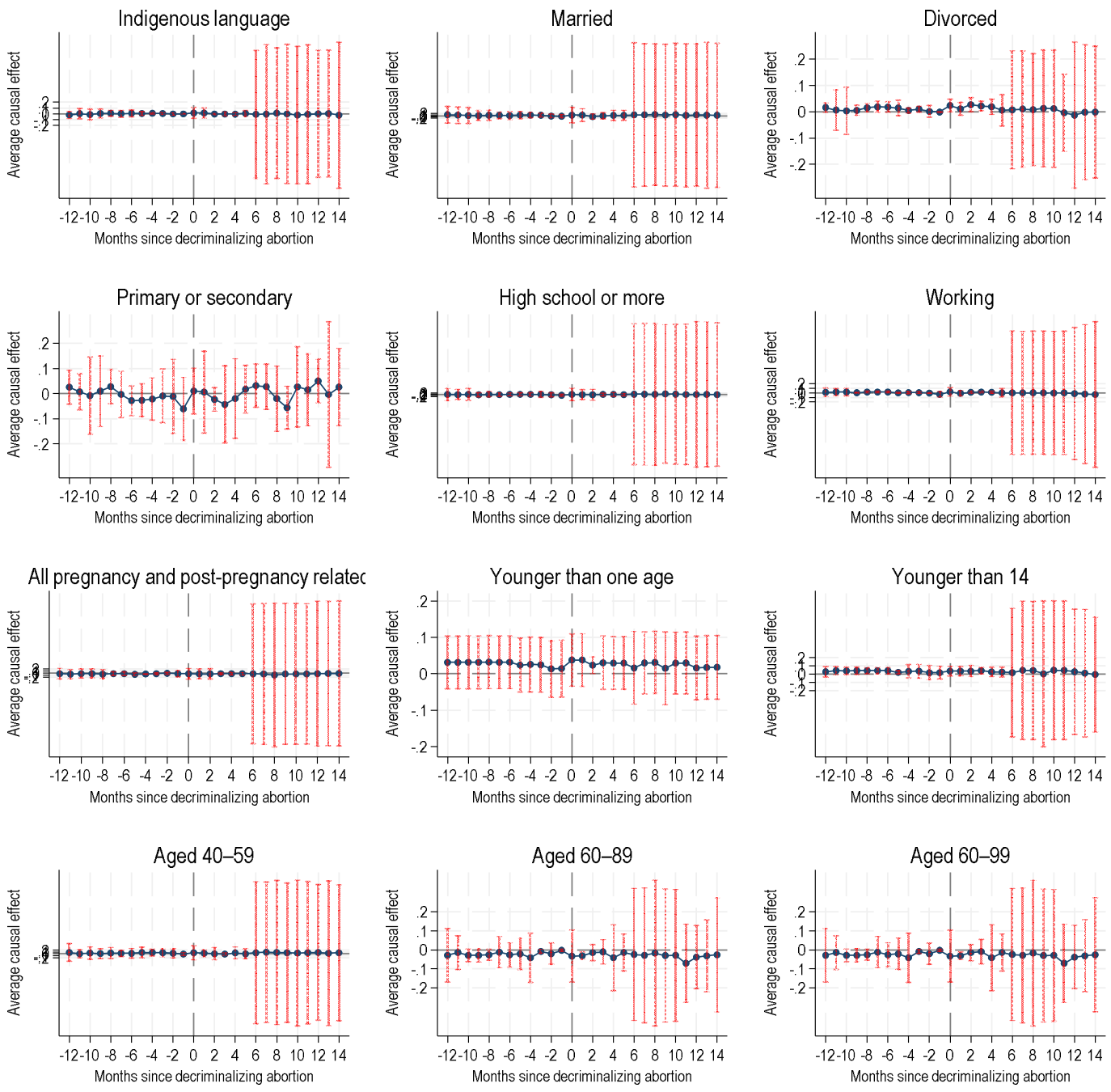

**Fig A.7.** The difference-in-differences of the decriminalization of abortion on homicides of women by subgroups

## Impact of harsher femicide penalties

**Table A.6**

Impact of the average prison sentences for femicides among states that implemented femicide laws without unilateral divorce laws or abortion decriminalization

| | Number of femicides | Number of homicides of women | Number of women who disappeared | Number of suicides of women |
|---|---|---|---|---|
| | (1) | (2) | (3) | (4) |
| DID estimate | 0.001 | 0.000 | 0.001 | -0.001 |
| | (0.002) | (0.002) | (0.001) | (0.002) |
| Number of observations x periods | 28582 | 28582 | 28582 | 28582 |
| Number of switchers x periods | 8117 | 8117 | 8117 | 8117 |
| Number of months used for post-treatment | 90 | 90 | 90 | 90 |
| Controls: unemployment rates of males and females, the number of males killed by firearms, area and fixed effects | Yes | Yes | Yes | Yes |
| Treatment | Enacting femicide law only (no unilateral divorce or decriminalization of abortion) | | | |
| Comparison | Without femicide laws, unilateral divorce or abortion decriminalization | | | |

*Notes*: Standard errors, clustered at the state level and bootstrapped, in parentheses. Significance levels *** p<0.01, ** p<0.05, * p<0.1.

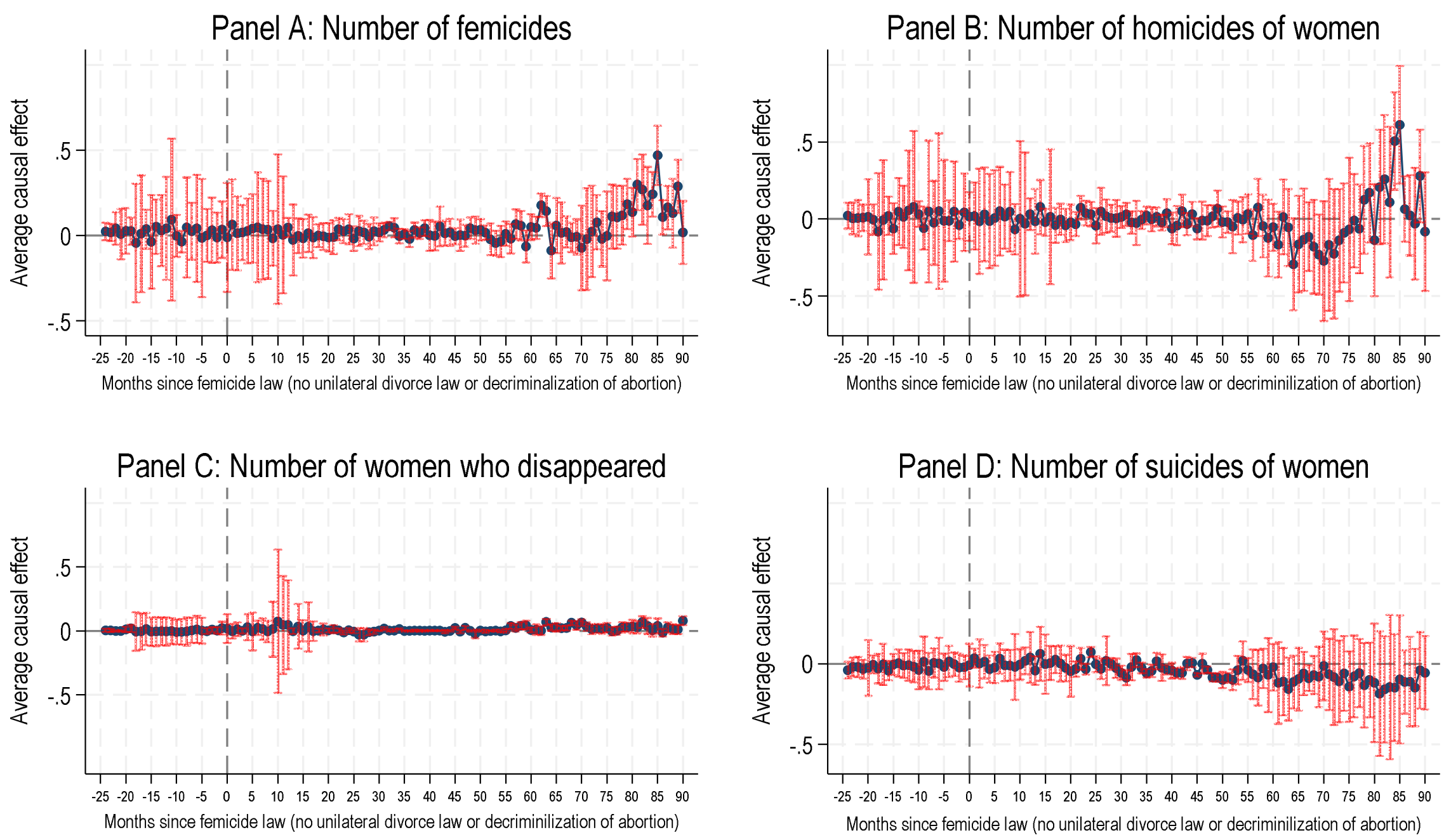

**Fig. A.8.** Impact of the average prison sentence for femicides among states that implemented femicide laws without unilateral divorce laws or abortion decriminalization

**Table A.7**

Impact of the average prison sentence for femicides among states that implemented femicide laws, regardless of whether they implemented other laws

| | Number of femicides | Number of homicides of women | Number of women who disappeared | Number of suicides of women |
|---|---|---|---|---|
| | (1) | (2) | (3) | (4) |
| DID estimate | 0.000 | 0.001 | 0.000 | 0.000 |
| | (0.000) | (0.001) | (0.000) | (0.000) |
| Number of observations x periods | 582866 | 582866 | 582866 | 582866 |
| Number of switchers x periods | 51741 | 51741 | 51741 | 51741 |
| Number of months used for post-treatment | 90 | 90 | 90 | 90 |
| Controls: unemployment rates of males and females, the number of males killed by firearms, area and fixed effects | Yes | Yes | Yes | Yes |
| Treatment | Enacting a femicide law and may have unilateral divorce, equitable divorce asset compensation or decriminalization of abortion | | | |
| Comparison | Without femicide law yet | | | |

*Notes*: Standard errors, clustered at the state level and bootstrapped, in parentheses. Significance levels *** p<0.01, ** p<0.05, * p<0.1.

**Table A.8**

Impact of the disparity in average prison sentences between femicides and homicides, among states that implemented femicide laws, regardless of whether they implemented other laws

| | Number of femicides | Number of homicides of women | Number of women who disappeared | Number of suicides of women |
|---|---|---|---|---|
| | (1) | (2) | (3) | (4) |
| DID estimate | 0.001 | 0.000 | -0.009 | 0.000 |
| | (0.002) | (0.003) | (0.006) | (0.002) |
| Number of observations x periods | 1129473 | 1129473 | 1129473 | 1129473 |
| Number of switchers x periods | 194085 | 194085 | 194085 | 194085 |
| Number of months used for post-treatment | 90 | 90 | 90 | 90 |
| Controls: unemployment rates of males and females, the number of males killed by firearms, area and fixed effects | Yes | Yes | Yes | Yes |
| Treatment | Enacting a femicide law and may have unilateral divorce, equitable divorce asset compensation or decriminalization of abortion | | | |
| Comparison | Without femicide law yet | | | |

*Notes*: Standard errors, clustered at the state level and bootstrapped, in parentheses. Significance levels *** p<0.01, ** p<0.05, * p<0.1.

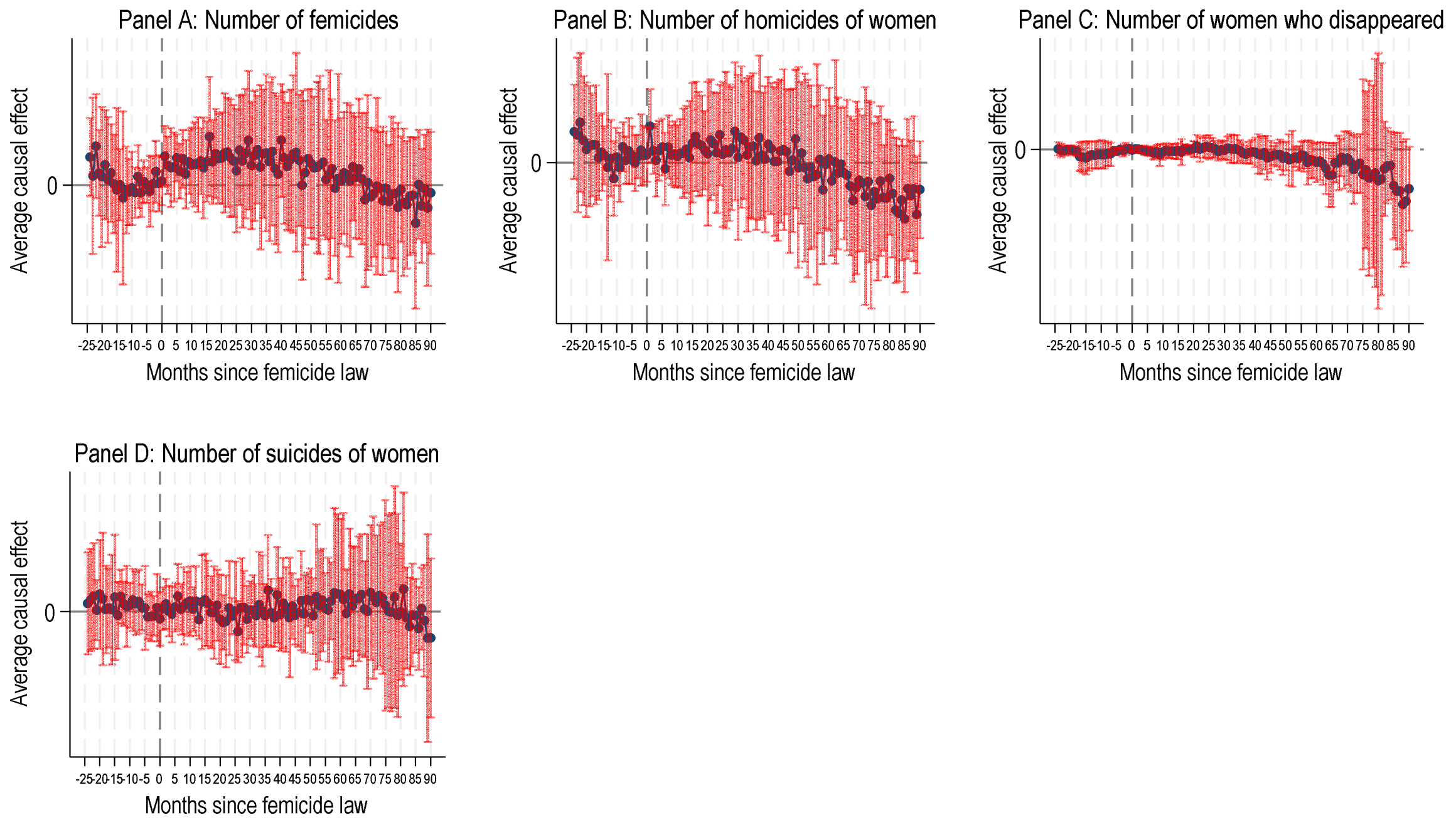

**Fig. A.9**. Impact of the disparity in average prison sentences between femicides and homicides, among states that implemented femicide laws, regardless of whether they implemented other laws

**Table A.9**

Impact of the average prison sentences for femicides on killing rates, among states that implemented femicide laws, regardless of whether they implemented other laws

| | Rate of femicides<br>(1) | Rate of homicides of women<br>(2) | Rate of women who disappeared<br>(3) | Rate of suicides of women<br>(4) |
|---|---|---|---|---|
| DID estimate | 0.001 | 0.001 | 0.002 | 0.000 |
| | (0.001) | (0.001) | (0.002) | (0.002) |
| Number of observations x periods | 580627 | 580627 | 580627 | 580627 |
| Number of switchers x periods | 51452 | 51452 | 51452 | 51452 |
| Number of months used for post-treatment | 90 | 90 | 90 | 90 |
| Controls: unemployment rates of males and females, the number of males killed by firearms, area and fixed effects | Yes | Yes | Yes | Yes |
| Treatment | Enacting a femicide law and may have unilateral divorce, equitable divorce asset | | | |
| Comparison | Without femicide law yet | | | |

*Notes*: Standard errors, clustered at the state level and bootstrapped, in parentheses. Significance levels *** p<0.01, ** p<0.05, * p<0.1.

**Table A.10**

Impact of the disparity in average prison sentences between femicides and homicides on killing rates, regardless of whether they implemented other laws

| | Rate of femicides<br>(1) | Rate of homicides of women<br>(2) | Rate of women who disappeared<br>(3) | Rate of suicides of women<br>(4) |
|---|---|---|---|---|
| DID estimate | 0.000 | 0.000 | -0.008 | 0.010 |
| | (0.001) | (0.001) | (0.009) | (0.007) |
| Number of observations x periods | 1124172 | 1124172 | 1124172 | 1124172 |
| Number of switchers x periods | 193144 | 193144 | 193144 | 193144 |
| Number of months used for post-treatment | 90 | 90 | 90 | 90 |
| Controls: unemployment rates of males and females, the number of males killed by firearms, area and fixed effects | Yes | Yes | Yes | Yes |
| Treatment | Enacting a femicide law and may have unilateral divorce or decriminalization of | | | |
| Comparison | Without femicide law yet | | | |

*Notes*: Standard errors, clustered at the state level and bootstrapped, in parentheses. Significance levels *** p<0.01, ** p<0.05, * p<0.1.

## MORE PLACEBO TESTS

**Table A.11**

Longer placebo tests: impact of femicide laws among states that implemented femicide laws, regardless of whether they implemented other laws

| | Number of femicides<br>(1) | Number of homicides of women<br>(2) | Number of women who disappeared<br>(3) | Number of homicides, disappearances, and suicides of women<br>(4) |
|---|---|---|---|---|
| DID estimate | 0.019 | 0.026 | -0.002 | 0.018 |
| | (0.012) | (0.021) | (0.012) | (0.037) |
| Number of observations x periods | 582866 | 582866 | 582866 | 582866 |
| Number of switchers x periods | 51741 | 51741 | 51741 | 51741 |
| Number of months used for post-treatment | 90 | 90 | 90 | 90 |
| Controls: unemployment rates of males and females, the number of males killed by firearms, area and fixed effects | Yes | Yes | Yes | Yes |
| Treatment | Enacted femicide law (regardless of whether it also has unilateral divorce, asset compensation, or decriminalization of abortion) | | | |
| Comparison | Without femicide law (regardless of whether it may have unilateral divorce, asset compensation, or decriminalization of abortion) | | | |

*Notes*: Standard errors, clustered at the state level and bootstrapped, in parentheses. Significance levels *** p<0.01, ** p<0.05, * p<0.1.

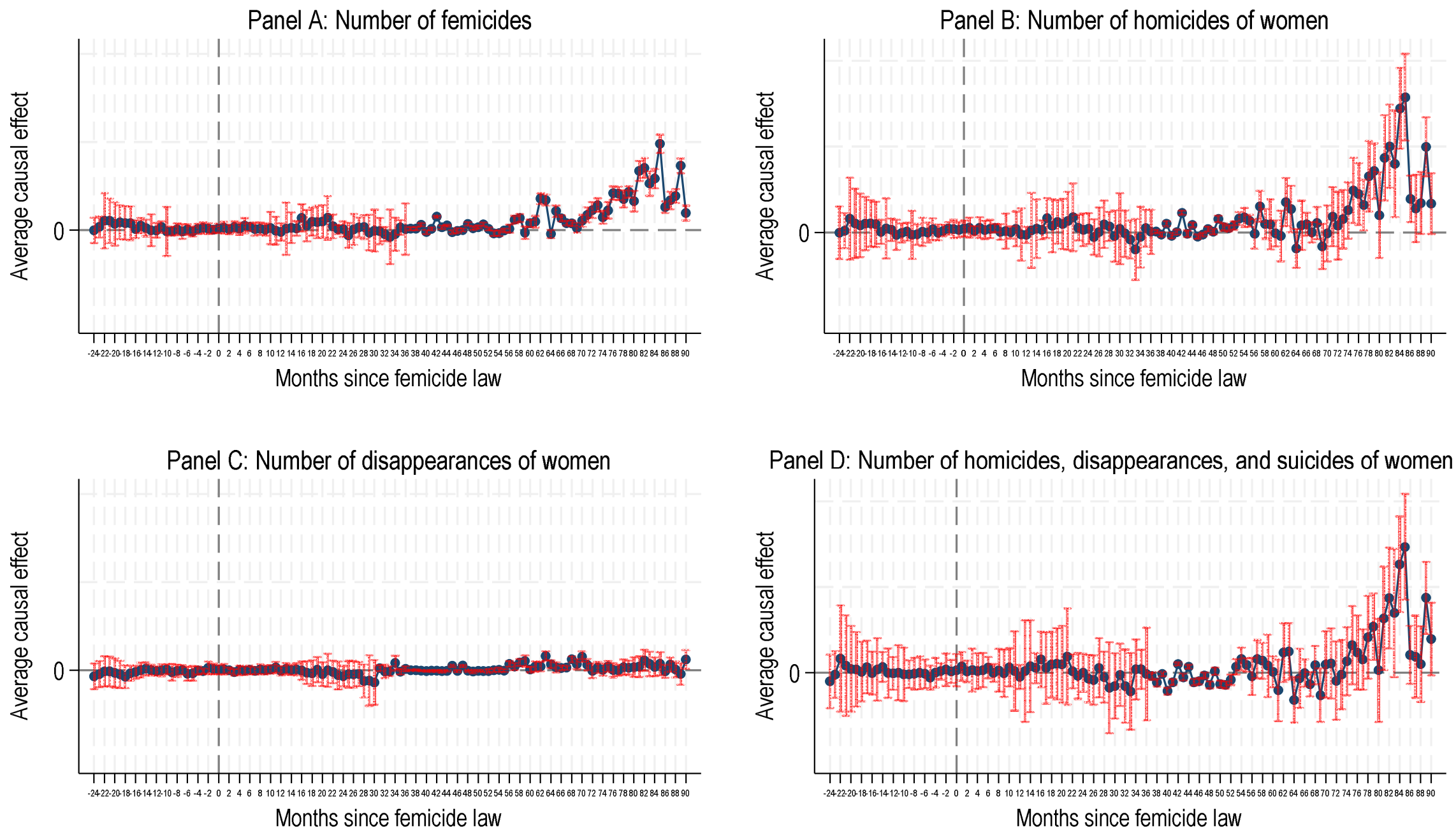

**Fig. A.10.** Longer placebo tests: Impact of femicide laws among states that implemented femicide laws, regardless of whether they implemented other laws

## Table A.12

Longer placebo tests: Alternative specification of impact of femicide laws among states that implemented femicide laws, regardless of whether they implemented other laws

| | Number of femicides | Number of homicides of women | Number of women who disappeared | Number of homicides, disappearances, and suicides of |
|---|---|---|---|---|
| | (1) | (2) | (3) | (4) |
| DID estimate | 0.019 | 0.025 | -0.004 | 0.019 |
| Standard error | (0.014) | (0.020) | (0.012) | (0.037) |
| Number of observations x periods | 582866 | 582866 | 582866 | 582866 |
| Number of switchers x periods | 51741 | 51741 | 51741 | 51741 |
| Number of months used for post-treatment | 90 | 90 | 90 | 90 |
| Controls: dummies for unilateral divorce, decriminalization of abortion, asset compensation after divorce, unemployment rates of males and females, the number of males killed by firearms, area, and fixed effects | Yes | Yes | Yes | Yes |
| Treatment | Enacted femicide law (regardless of whether it also has unilateral divorce, asset compensation, or decriminalization of abortion) | | | |
| Comparison | Without femicide law (regardless of whether it may have unilateral divorce, asset compensation, or decriminalization of abortion) | | | |

*Notes*: Standard errors, clustered at the state level and bootstrapped, in parentheses. Significance levels *** p<0.01, ** p<0.05, * p<0.1.

## Divorce asset compensation

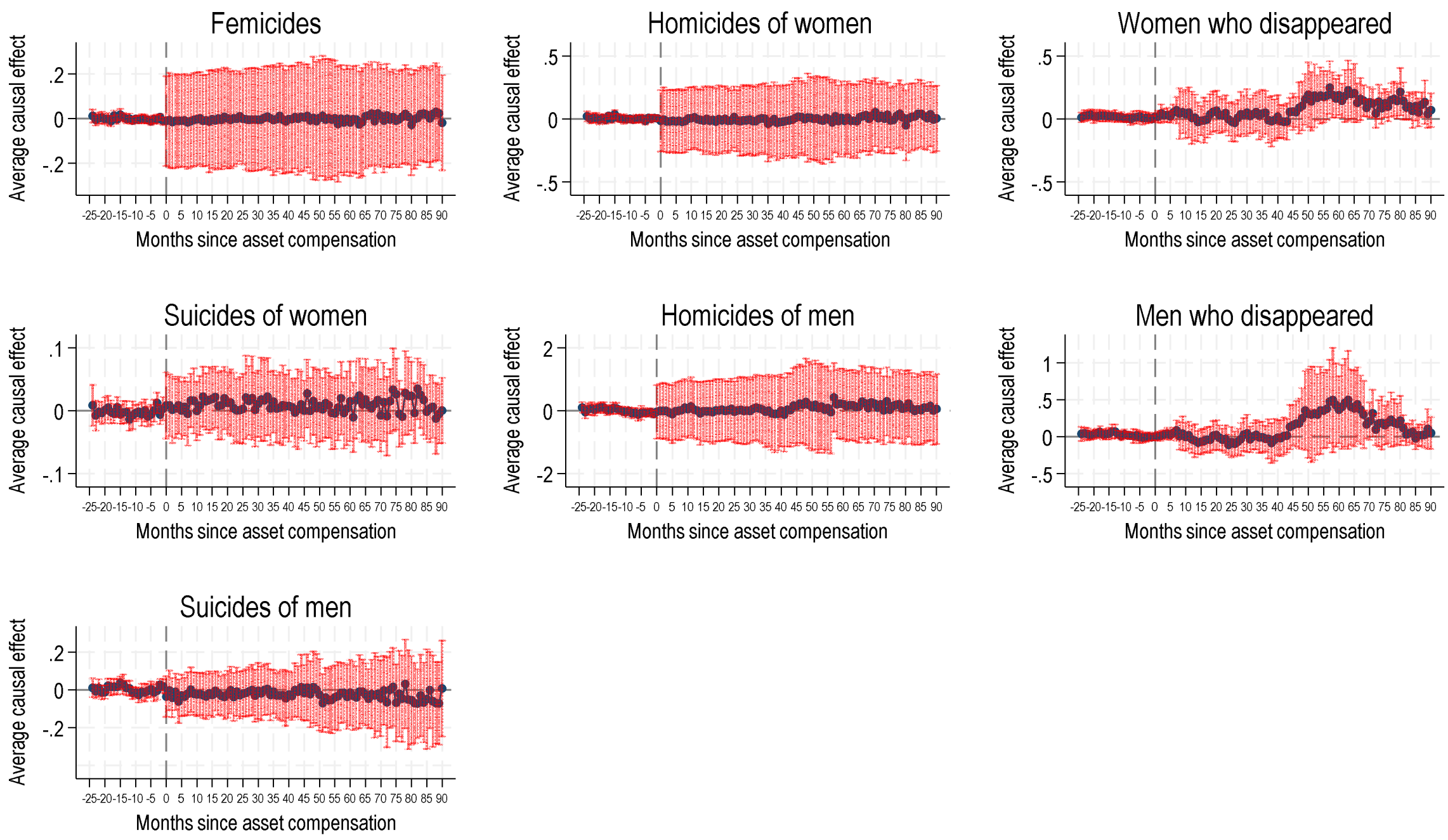

**Fig. A.11.** Impact of the equitable divorce asset compensation law on lethal violence

**Table A.13**

Impact of enacting the equitable divorce asset compensation law on lethal violence

| | Number of femicides (1) | Number of homicides of women (2) | Number of women who disappeared (3) | Number of suicides of women (4) | Number of homicides of men (5) | Number of men who disappeared (6) | Number of suicides of men (7) |
|---|---|---|---|---|---|---|---|
| DID estimate | -0.003 | -0.007 | 0.054 | 0.009 | 0.039 | 0.065 | -0.023 |
| | (0.114) | (0.14) | (0.049) | (0.025) | (0.515) | (0.082) | (0.063) |
| Number of observations x periods | 2294905 | 2294905 | 2294905 | 2294905 | 2294905 | 2294905 | 2294905 |
| Number of switchers x periods | 105557 | 105557 | 105557 | 105557 | 105557 | 105557 | 105557 |
| Number of months used for post-treatment | 90 | 90 | 90 | 90 | 90 | 90 | 90 |
| Controls: area, time fixed effects, along with two dummy variables indicating whether the state has implemented unilateral and femicide laws | Yes | Yes | Yes | Yes | Yes | Yes | Yes |
| Treatment | Enacted the equitable divorce asset compensation law (regardless of whether it may have femicide, unilateral divorce, or decriminalized abortion) | | | | | | |
| Comparison | Without equitable divorce asset compensation law | | | | | | |

*Notes*: Standard errors, clustered at the state level and bootstrapped, in parentheses.

Significance levels *** p<0.01, ** p<0.05, * p<0.1.

**Table A.14**

Impact of enacting both the divorce asset compensation and the unilateral divorce laws

| | Number of femicides (1) | Number of homicides of women (2) | Number of women who disappeared (3) | Number of suicides of women (4) | Number of homicides of men (5) | Number of men who disappeared (6) | Number of suicides of men (7) |
|---|---|---|---|---|---|---|---|
| DID estimate | -0.135 | -0.213 | 0.087 | 0.008 | -2.160 | -0.134 | -0.016 |
| | (0.161) | (0.217) | (0.064) | (0.037) | (2.856) | (0.154) | (0.091) |
| Number of observations x periods | 60714 | 60714 | 60714 | 60714 | 60714 | 60714 | 60714 |
| Number of switchers x periods | 26368 | 26368 | 26368 | 26368 | 26368 | 26368 | 26368 |
| Number of months used for post-treatment | 90 | 90 | 90 | 90 | 90 | 90 | 90 |
| Controls: area, time fixed effects, along with a dummy variable indicating whether the state has implemented femicide laws | Yes | Yes | Yes | Yes | Yes | Yes | Yes |
| Treatment | Enacted both the equitable divorce asset compensation and the unilateral divorce law | | | | | | |
| Comparison | Enacted equitable divorce asset compensation law but has not implemented the unilateral divorce law yet | | | | | | |

*Notes*: Standard errors, clustered at the state level and bootstrapped, in parentheses. Significance levels *** p<0.01, ** p<0.05, * p<0.1.

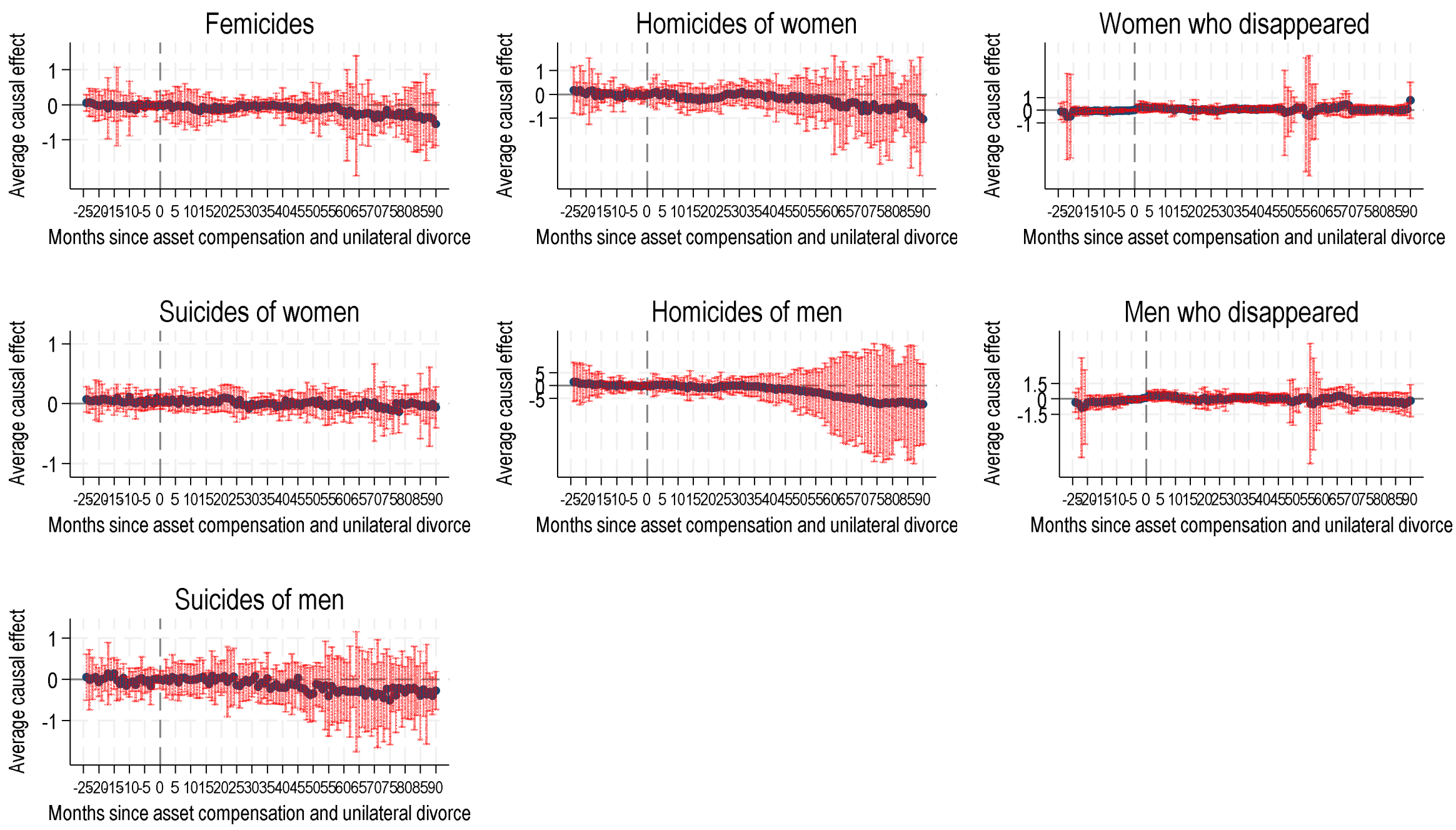

**Fig. A.12.** Impact of enacting both the divorce asset compensation and the unilateral divorce laws

## Table A.15

Impact of enacting the equitable divorce asset compensation law by marital status

| | Number of homicides of married women (1) | Number of homicides of divorced women (2) | Number of homicides of separated women (3) | Number of homicides of married men (4) | Number of homicides of divorced men (5) | Number of homicides of separated men (6) |
|---|---|---|---|---|---|---|
| DID estimate | 0.004 | 0.000 | -0.003 | 0.061 | -0.001 | 0.005 |
| | (0.016) | (0.003) | (0.022) | (0.182) | (0.011) | (0.022) |
| Number of observations x periods | 2294905 | 2294905 | 2294905 | 2294905 | 2294905 | 2294905 |
| Number of switchers x periods | 105557 | 105557 | 105557 | 105557 | 105557 | 105557 |
| Number of months used for post-treatment | 90 | 90 | 90 | 90 | 90 | 90 |
| Controls: area, time fixed effects, along with two dummy variables indicating whether the state has implemented unilateral and femicide laws | Yes | Yes | Yes | Yes | Yes | Yes |
| Treatment | Enacted the equitable divorce asset compensation law (regardless of whether it may have femicide, unilateral divorce, or decriminalized abortion) | | | | | |
| Comparison | Without equitable divorce asset compensation law | | | | | |

*Notes*: Standard errors, clustered at the state level and bootstrapped, in parentheses.

Significance levels *** p<0.01, ** p<0.05, * p<0.1.

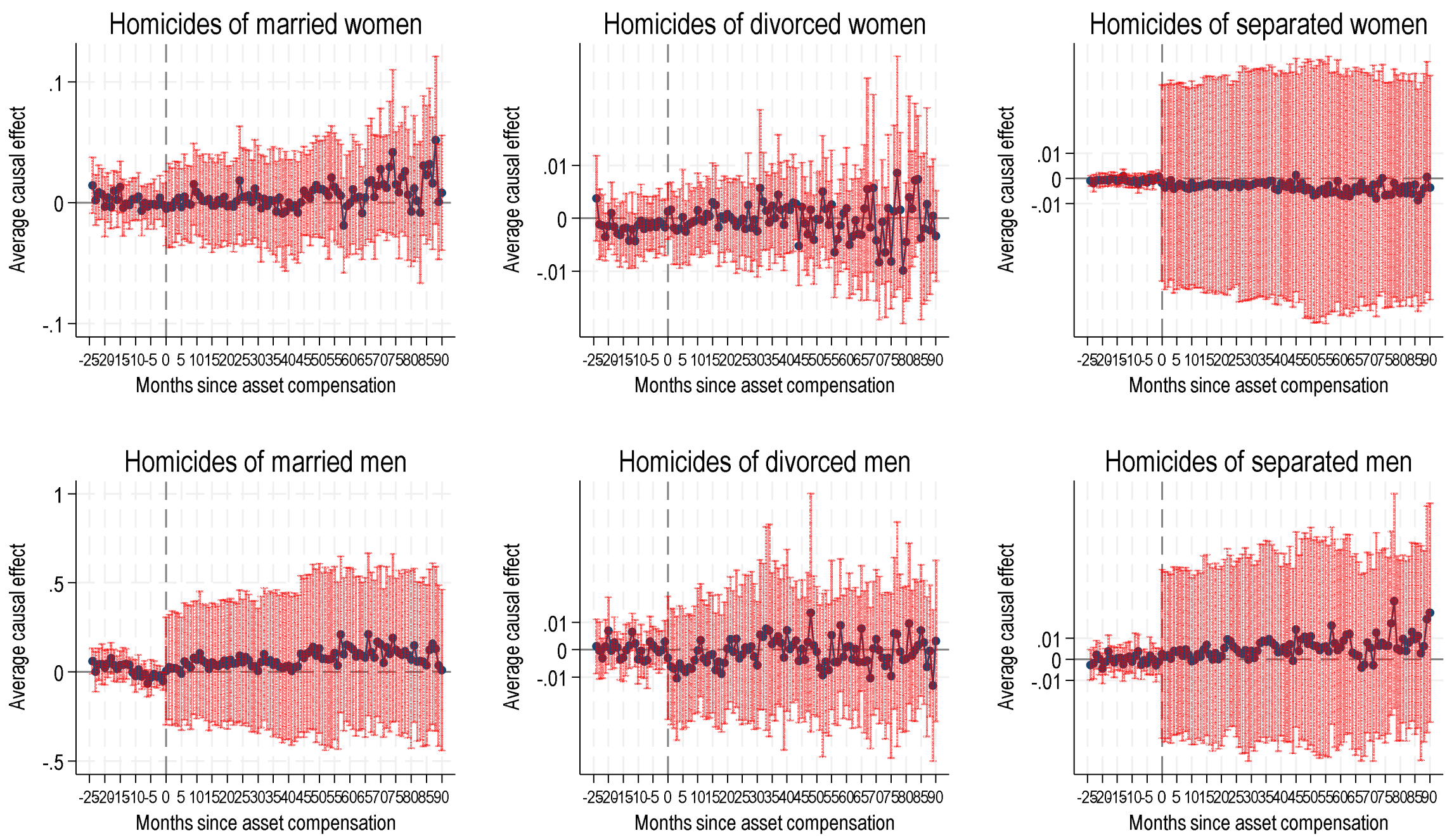

**Fig. A.13.** Impact of enacting the equitable divorce asset compensation law by marital status

**Table A.16**

Impact of enacting the equitable divorce asset compensation and the unilateral divorce laws by marital status

| | Number of homicides of married women | Number of homicides of divorced women | Number of homicides of separated women | Number of homicides of married men | Number of homicides of divorced men | Number of homicides of separated men |
|---|---|---|---|---|---|---|
| | (1) | (2) | (3) | (4) | (5) | (6) |
| DID estimate | -0.033 | -0.002 | -0.003 | -0.804 | 0.006 | -0.036 |
| | (0.057) | (0.022) | (0.010) | (0.485) | (0.036) | (0.048) |
| Number of observations x periods | 60714 | 60714 | 60714 | 60714 | 60714 | 60714 |
| Number of switchers x periods | 26368 | 26368 | 26368 | 26368 | 26368 | 26368 |
| Number of months used for post-treatment | 90 | 90 | 90 | 90 | 90 | 90 |
| Controls: area, time fixed effects, along with a dummy variable indicating whether the state has implemented femicide laws | Yes | Yes | Yes | Yes | Yes | Yes |
| Treatment | Enacted both the equitable divorce asset compensation and the unilateral divorce law | | | | | |
| Comparison | Enacted equitable divorce asset compensation law but has not implemented the unilateral divorce law yet | | | | | |

*Notes*: Standard errors, clustered at the state level and bootstrapped, in parentheses. Significance levels *** p<0.01, ** p<0.05, * p<0.1.

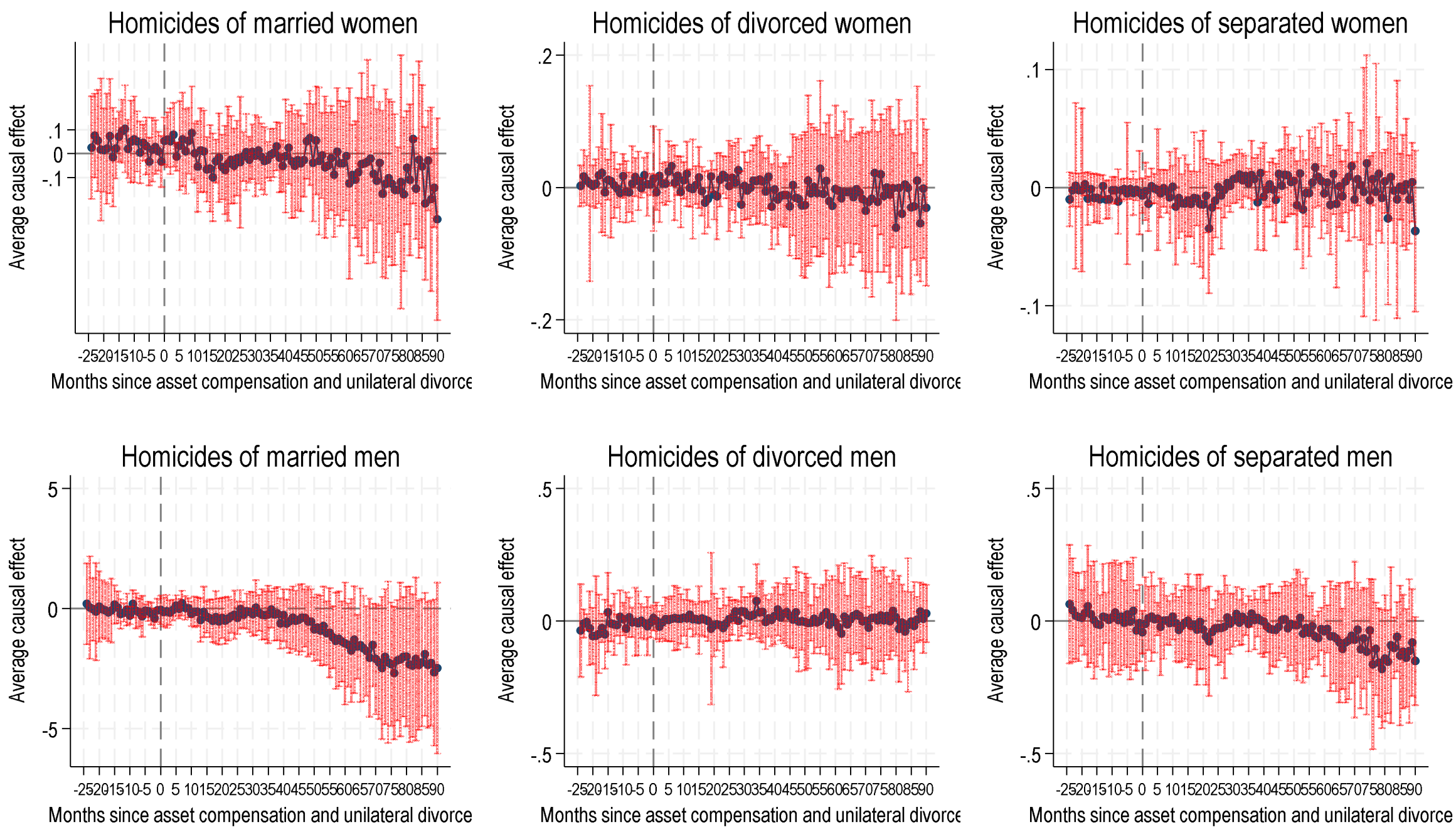

**Fig. A.14**. Impact of enacting the divorce asset compensation and the unilateral divorce laws by marital status